\documentclass{egpubl}
\usepackage{eg2024}
 
\ConferencePaper        
\CGFccby

\usepackage[T1]{fontenc}
\usepackage{dfadobe}  

\usepackage{cite}  
\BibtexOrBiblatex
\electronicVersion
\PrintedOrElectronic
\ifpdf \usepackage[pdftex]{graphicx} \pdfcompresslevel=9
\else \usepackage[dvips]{graphicx} \fi

\usepackage{egweblnk} 

\usepackage{amsmath}
\usepackage{amssymb}
\usepackage{array}
\usepackage[labelfont=bf, font=it]{caption}
\usepackage{graphicx}
\usepackage{multirow}
\usepackage{pgfplots}
\pgfplotsset{compat=1.15}
\usepackage{pifont}
\usepackage{threeparttablex}

\title[Strokes2Surface]%
      {Strokes2Surface: Recovering Curve Networks From 4D Architectural Design Sketches}

\author[S. Rasoulzadeh, M. Wimmer, P. Stauss, I. Kovacic]{
    \parbox{\textwidth} {
            \centering 
            S. Rasoulzadeh\orcid{0000-0002-0019-0137}, 
            M. Wimmer\orcid{0000-0002-9370-2663}, 
            P. Stauss\orcid{0009-0004-0573-5441}, 
            I. Kovacic\orcid{0000-0002-0303-3284}
        }\\
        \parbox{\textwidth} {
            \centering 
            TU Wien, Center for Geometry and Computational Design, Austria
        }
}

\begin{document}

\teaser{
    \includegraphics[width=1.0\linewidth]{FIGURES/00_TEASER.png}
    \centering
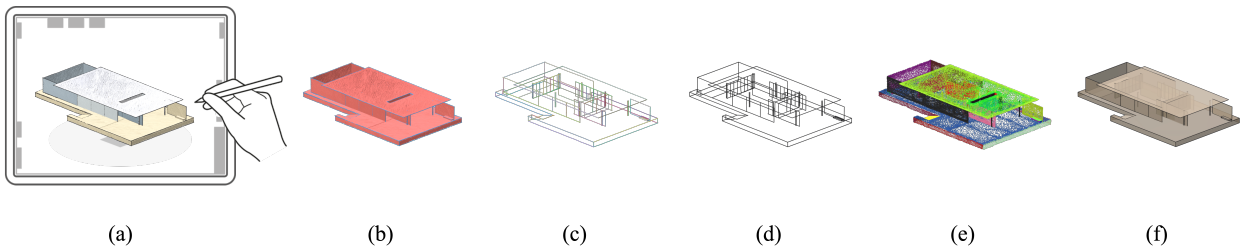
    \caption{Overview of the Strokes2Surface pipeline. (a) The designer creates a sketch of an architectural object. (b) The pipeline’s binary classifier distinguishes between Shape strokes depicting edges (blue) and Scribble strokes depicting faces (red). (c) Shape strokes are parsed into clusters and consolidated representing single edges, (d) the intended topology is recovered, forming a well-connected curve network, (e) Scribble strokes are parsed into clusters representing faces of the object corresponding to curve network cycles, and (f) the final reconstructed surface mesh. Images rendered using Polyscope \cite{polyscope}.}
    \label{fig:TEASER}
}

\maketitle
\begin{abstract}    
    We present Strokes2Surface, an offline geometry reconstruction pipeline that recovers well-connected curve networks from imprecise 4D sketches to bridge concept design and digital modeling stages in architectural design. The input to our pipeline consists of 3D strokes' polyline vertices and their timestamps as the 4th dimension, along with additional metadata recorded throughout sketching. Inspired by architectural sketching practices, our pipeline combines a classifier and two clustering models to achieve its goal. First, with a set of extracted hand-engineered features from the sketch, the classifier recognizes the type of individual strokes between those depicting boundaries (Shape strokes) and those depicting enclosed areas (Scribble strokes). Next, the two clustering models parse strokes of each type into distinct groups, each representing an individual edge or face of the intended architectural object. Curve networks are then formed through topology recovery of consolidated Shape clusters and surfaced using Scribble clusters guiding the cycle discovery. Our evaluation is threefold: We confirm the usability of the Strokes2Surface pipeline in architectural design use cases via a user study, we validate our choice of features via statistical analysis and ablation studies on our collected dataset, and we compare our outputs against a range of reconstructions computed using alternative methods. \\

\ccsdesc[300]{Computing methodologies~Artificial intelligence}
\ccsdesc[300]{Computing methodologies~Computer graphics}
\ccsdesc[300]{Computing methodologies~Machine learning}

\printccsdesc   
\end{abstract}  

\section{Introduction}
\label{sec:INTRODUCTION}

Freeform architectural design often involves two essential stages: concept design and digital modeling \cite{deng2022sketch2pq}. During the former stage, designers typically favor freehand sketching due to its low overhead in representing, exploring, and communicating geometric ideas \cite{chen2008sketching}. In this stage, sketching allows designers to easily tap into their intuition and enter a state of mental flow that otherwise might be difficult to achieve \cite{mahoney2018v}. Subsequent to the concept design, the digital modeling stage ensues. In this stage, taking the sketch as a visual reference, the architect or designer usually uses 3D modeling software to manually create and edit a 3D digital model into the desired shape, which can be subsequently processed for presentation, structural analysis, manufacturing, or other downstream design pipelines. 

Traditionally, architects and designers relied on pen and paper as their primary drawing medium during the ideation process in the concept design stage. However, with the growing availability of 3D sketching interfaces and Augmented, Virtual, and Mixed Reality (AR/VR/MR) technologies \cite{dorsey2007mental, mahoney2018v, zheng2016smartcanvas, xin2008napkin}, a paradigm shift is being witnessed. These emerging environments and technologies allow designers to sketch in 3D and, in turn, enhance efficiency and maintain better supervision over their creations. This shift has piqued the attention of architects and experts, sparking explorations of the potential of architectural design within these environments \cite{dzurillas}. When it comes to the digital modeling stage, manually translating a sketch into a digital model often requires a considerable amount of time and is susceptible to misinterpretations, underscoring the need for an automated reconstruction pipeline. Recent studies \cite{xu2014true2form, gryaditskaya2020lifting, hahnlein2022symmetry, tono2022vitruvio} have shown that 2D design sketches can be lifted into 3D sketch or 3D digital models. Also, several 3D, AR, VR, and MR academic and commercial interfaces come with a coupled reconstruction pipeline allowing sketch-based modeling \cite{rosales2019surfacebrush, yu2021cassie}. However, most of these systems primarily target industrial design, which functions under its own assumptions that may not necessarily align with architectural design, putting a question on their applicability for such purposes. More specifically, many existing 3D sketch-based modeling methods often either impose certain interface-specific constraints that designers must adhere to or progressively neaten the designer-drawn strokes on the fly in an \textit{online} manner to facilitate the modeling. Such constraints and interaction setups can inhibit the designer's freedom and disrupt the ideation process, leading architects to feel like losing authorship and control over their creations \cite{do2002drawing}. The noted barriers highlight the need for an automated geometry reconstruction method, tailored specifically for architectural design, functioning akin to a "magical button", seamlessly capturing the design intention and translating the \textbf{completed} sketch into the desired geometry in an \textit{offline} manner.

To this end, building upon \textit{MR.Sketch}, a pen-on-tablet 4D sketching interface targeted for architectural design developed within our research group \cite{kovacs2023mr}, we introduce the \textit{Strokes2Surface} geometry reconstruction pipeline. Strokes2Surface processes 4D architectural design sketches to form well-connected curve networks, thereby bridging the gap between the concept design and digital modeling stages. The input to our pipeline consists of the 4D sketch strokes, along with the additional metadata (geometry and stylus-related properties) actively recorded at the time of sketching. Given this, our pipeline is motivated by architectural sketching practices, the inherent characteristics of the inputs, and our thorough analysis of the recorded metadata:
\begin{itemize}
    \item Our studies and observations of architectural sketching practices reveal that designers draw two types of strokes when depicting their geometric ideas. One set of strokes outlines the boundaries and edges of the intended geometry, while the other type is drawn to fill in and mark mark the faces of the intended geometry (Section \ref{sec:INPUT_DRAWING_CHARACTERISTICS}). 
    \item Concept design sketches are often imprecise, exhibiting over-sketching or gaps and missing intersections between strokes.
    \item Our thorough analysis shows that the fourth dimension, along with the recorded metadata during sketching, inherently carries valuable information pertaining to the design intent. (Sections \ref{sub_sub_sec:FEATURE_EXTRACTION}, \ref{sub_sub_sec:FEATURE_SIGNIFICANCE_TESTING}, and \ref{sub_sub_sec:ABLATION STUDIES}).
\end{itemize}
In summary, Strokes2Surface is an offline pipeline comprising three Machine Learning (ML) models. The first model is a meta estimator, which is responsible for classifying the type of drawn strokes (Section \ref{sub_sec:STROKE_TYPE_RECOGNITION}) trained using a curated dataset of architectural design sketches (Section \ref{sub_sec:DATASET}). Then, there are two clustering models that further parse strokes of each type into separate groups representing a set of potentially over-sketched strokes forming either a single boundary and edge, or a single face of the geometry, respectively (Sections \ref{sub_sub_sec:CLUSTERING_SHAPE_STROKES} and \ref{sub_sub_sec:CLUSTERING_SCRIBBLE_STROKES}). The groups resulting from the first clustering model are further consolidated into 3D aggregate curves by using cubic B-spline approximation, and a well-connected curve network is then formed by recovering the intended topology of the curves by formulating it as a minimization problem (Section \ref{sub_sub_sec:TOPOLOGY_RECOVERY}). Furthermore, the groups obtained by the second clustering model guide the pipeline to infer which curve network cycles must bound patches and which must not, ultimately facilitating the reconstruction of the user-intended geometry (Section \ref{sub_sub_sec:CYCLE DISCOVERY_AND_GEOMETRY_GENERATION}) (Figure \ref{fig:TEASER}). Our code and data are available at: \URL{https://gitlab.cg.tuwien.ac.at/srasoulzadeh/strokes2surface.git}.

Our evaluation of the pipeline is threefold: We confirm our studies of architectural sketching practices and the usability of the Strokes2Surface pipeline in architectural design use cases via a user study (Section \ref{sec:USER_STUDY}), we validate our choice of features via statistical analysis and ablation tests on our collected dataset (Section \ref{sub_sub_sec:FEATURE_SIGNIFICANCE_TESTING} and \ref{sub_sub_sec:ABLATION STUDIES}), and we compare the outputs produced by different steps of the pipeline against point cloud, stroke cloud, and curve network surfacing methods (Section \ref{sub_sub_sec:COMPARISON_TO_PRIOR_ART}). 


\section{Related Work}
\label{sec:RELATED_WORKS}

Our work builds upon prior research across multiple domains.

\paragraph*{Sketch Consolidation.} When creating sketches, designers frequently depict their intended curves using multiple, tightly clustered, or over-sketched strokes. Sketch beautification and consolidation methods parse such strokes into groups that jointly define intended aggregate curves. \cite{barla2005geometric} was among the first to present an algorithm for line drawing simplification, in which a complex vector graphic drawing is "redrawn" with fewer strokes. They address this problem by clustering strokes and then replacing each cluster with a single representative curve. Thereafter, numerous works followed this same basic cluster-and-replace framework \cite{orbay2011beautification, ogawa2016sketch, liu2018strokeaggregator, van2021strokestrip}. \cite{chien2014line} applies a low-pass Gaussian filter to translate the strokes based on the weight of the filtering, and then the strokes are paired based on the strokes' endpoints' positions and tangents. In terms of learning-based methods, \cite{orbay2011beautification} tackles this problem by using a neural network taking as input a set of geometric features extracted from each pair of strokes and classifying whether the two strokes should reside in the same group or not. Similarly, \cite{ogawa2016sketch} trained a support vector machine to estimate the pair of strokes to be merged. Significantly improving on the state-of-the-art are the two recent works StrokeAggregator \cite{liu2018strokeaggregator} and StrokeStrip \cite{van2021strokestrip}. \cite{liu2018strokeaggregator} uses human perception principles and artistic practice observations for stroke clustering, employing angular and proximity scores with HDBSCAN for incremental merging, followed by local cluster refinement and curve fitting. Given a vector sketch with multiple overdrawn strokes, \cite{liu2023stripmaker} consolidates it through two classifiers: the first evaluates strokes locally, while the second, incorporates global context, refining the first classifier's consolidation output. The second step in our pipeline shares similarities with existing 2D sketch consolidation methods. However, in 3D, some assumed properties in 2D sketches no longer hold (Section \ref{sub_sub_sec:CLUSTERING_SHAPE_STROKES}), requiring a tailored 3D method. A recent interface targeting industrial design, ScaffoldSketch \cite{yu2021scaffoldsketch}, facilitates in-air design drawing using a two-stage approach akin to 2D design. It decomposes strokes into Scaffold and Shape types, auto-correcting and beautifying them for aesthetic, accurate 3D drawings.

There are several 3D sketching interfaces that solely focus on novel interaction techniques, while some others are also coupled with modeling algorithms. Additionally, there are several standalone surfacing methods developed specifically for inputs originating from typical 3D sketching interfaces, such as curve networks or stroke clouds, or can be applied to other geometric formats produced from 3D sketches, such as point clouds. 

\paragraph*{3D Sketching Interfaces.} A prominent interface is MentalCanvas \cite{dorsey2007mental}, designed to allow architects to organize concept drawings in 3D by first making several regular 2D sketches of their design from different viewpoints and then fusing them together into a 3D structure. A number of other sketching systems follow 3D projective sketching \cite{xin2008napkin, lee2022rapid}. NapkinSketch \cite{xin2008napkin} allows users to draw 3D sketches on top of a drawing canvas set up by pen strokes. A more recent work of a similar vein is \cite{lee2022rapid}, where they present a pen-on-tablet system featuring multi-touch gestures developed for rapidly creating concepts of articulated objects. Among VR sketch-based modeling interfaces, \cite{mahoney2018v} introduced a prototype for a 3D sketching interface in architecture, utilizing machine learning to translate sketches into 3D forms through conversion to an intermediate description followed by the use of a reconstruction function. CASSIE \cite{yu2021cassie} is another VR-based conceptual modeling system that leverages freehand mid-air sketching coupled with a 3D optimization framework performing automatic surface neatening in real-time, resulting in well-connected 3D curve networks. 

\paragraph*{Surfacing Point Clouds.} 3D sketches often can be converted into point clouds by sampling points along each stroke, allowing leveraging the wealth of both non-data-driven and data-driven methods developed for surfacing point clouds. However, such samplings often produces very sparse point clouds, exhibiting inconsistent normal orientation, multiple samples in the interior of the intended object, and other artifacts inconsistent with the assumptions made by typical reconstruction techniques. In terms of early non-data-driven methods, VIPSS \cite{huang2019variational}, a method applicable to point sets obtained from 3D sketches, reconstructs implicit surfaces from un-oriented sets using quadratic optimization but struggles with sharp surface discontinuities common in architectural structures. Screened Poisson Reconstruction \cite{kazhdan2013screened} is the most commonly used method to convert an unstructured point cloud along with its per-point normals to a surface mesh. However, the absence of data-priors in these type of methods makes them fail to handle noisy inputs, which is a very common case in point clouds produced by sketches. Points2Surf \cite{erler2020points2surf} is a data-driven patch-based learning framework creating surfaces from raw scans without needing normals, trained on solid objects. Learning a prior over a combination of detailed local patches and coarse global information improves generalization performance and reconstruction accuracy. However, their network, faces challenges in reconstructing surfaces from point clouds of non-solid 3D sketched objects. Yet another recent interesting point cloud reconstruction method is Point2Mesh \cite{hanocka2020point2mesh} which optimizes the weights of a Convolutional Neural Network (CNN) to deform an initial mesh to shrink-wrap the input point cloud. The optimized CNN weights act as a prior, which encode the expected shape properties and converge to a desirable solution. In Section \ref{sub_sub_sec:COMPARISON_TO_PRIOR_ART}, we produce point clouds from our 3D sketches and compare our specialized reconstructions against three of point cloud reconstruction methods.

\paragraph*{Surfacing Stroke Clouds.} In the computer graphics and vision community, a few methods have been proposed to reconstruct 3D geometries from stroke clouds \cite{fabbri20103d, batuhan2012free, yu2022piecewise, luo20233d}. Relying on the images from coarsely calibrated cameras, \cite{fabbri20103d} developed a framework for 3D reconstruction from an unorganized set of curves. Close to our work is SurfaceBrush \cite{rosales2019surfacebrush}, which is also an offline reconstruction method for freeform surface modeling from input cloud of 3D stroke ribbons. Their specialized surfacing algorithm, coupled with their sketching interface, works by converting raw designer-drawn strokes into a user-intended manifold 3D surface by matching edge sequences along input stroke polylines. \cite{batuhan2012free} surfaces sparse and imprecise 3D sketches by smoothly deforming an initial low-fidelity surface of correct topology, using a discrete guidance vector field that points towards the closest stroke point. This approach produces globally smooth surfaces but requires user intervention to specify strokes that should be inserted into the mesh as sharp edge polylines. \cite{yu2022piecewise} transforms sparse 3D stroke clouds into piecewise-smooth surfaces using iterative segmentation and optimization of smooth patches to fit surrounding strokes. Unlike our pipeline, which requires no user intervention, both two latter methods rely on user-annotation to determine boundary strokes to trim the surface. Our inputs do not conform their input specifications but the output that the second step of our pipeline produces can be regarded as stroke clouds and fed to their method. We provide detailed comparison of our reconstructions to \cite{yu2022piecewise} in Section \ref{sub_sub_sec:COMPARISON_TO_PRIOR_ART}.

\paragraph*{Surfacing Curve Networks.} The advent of practical interfaces and devices has motivated the development of algorithms to automatically surface a sparse, designer-drawn set of well-connected curves, so-called curve networks. In general, surfacing curve networks is considered a two-step challenge. The first challenge is to discover, among all closed cycles in the curve network, which closed cycles should be delimiting surface patches and which should not \cite{abbasinejad2012surface, zhuang2013general}. A second challenge is to generate the surface geometry that interpolates the cycle boundaries by propagating geometric information on the boundary curves \cite{zou2013algorithm, pan2015flow, stanko2016smooth}. Finding the optimal set of cycles often corresponds to a \textit{cycle basis} in graph representation of the curve network. A cycle basis is a minimal set of cycles in a graph such that any cycle not in the basis can be constructed by the \textit{ring sum} of some cycles in the basis. The cycle basis of a graph can be computed easily using spanning trees, but the computed basis may contain cycles not corresponding to desired surface patches. Given the graph representation of the curve network, \cite{abbasinejad2012surface} addresses this problem by starting with an initial cycle basis and then uses a greedy algorithm to construct the optimal one. On the other hand, \cite{zhuang2013general} considers an alternative representation called a \textit{routing system}, which implicitly encodes a set of cycles by local variables at each vertex and each curve of the network. Optimizing cost metrics designed for these variables, they are able to compute cycles more efficiently and handle inputs with more complex topology and geometry. So as to solve the second challenge, \cite{zou2013algorithm} investigated an algorithm to obtain a triangulation of multiple and non-planar 3D polygons while minimizing additive weights, such as the total triangle areas or the total dihedral angles between adjacent triangles. \cite{stanko2016smooth} is yet another 3D curve network surfacing algorithm, which requires surface normals as well. The algorithm has been proposed to improve the quality of the generated surfaces and could be applied to our results as a post-processing. The outputs from the very last step of our proposed pipeline, can be benchmarked against these methods designed to tackle the two previously mentioned challenges: cycle finding and curve network surfacing. We compare our approach toward guiding the cycle discovery leveraging clustered Scribble strokes by comparing against state-of-the-art cycle identification method proposed in \cite{zhuang2013general} and use the method of \cite{zou2013algorithm} for triangulating the cycles.


\section{The 4D Drawing Interface}
\label{sec:THE_4D_DRAWING_INTERFACE}

Strokes2Surface is built upon a 4D sketching interface named MR.Sketch \cite{kovacs2023mr}, targeted for architectural design, utilizing a tablet (iPad) and a stylus (Apple Pencil) as its primary drawing medium. The interface enables the creation of 4D sketches by employing 3D canvases, where the 2D strokes drawn on the tablet's surface are projected onto the canvases, forming 3D strokes. Additionally, temporal data of all strokes is continuously captured throughout the drawing process, providing the 4th dimension, namely timestamp, to the sketch and its constituent strokes.

\begin{figure}[]
    \centering
    \includegraphics[width=.50\linewidth]{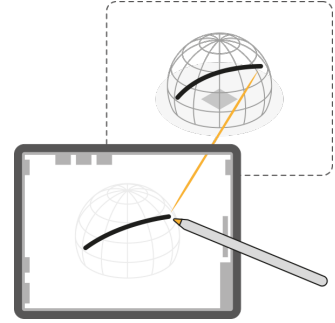}
    \caption{3D stroke formation through ray casting. As 2D strokes are drawn on the tablet's surface, rays originating from the camera's viewpoint are intersected the selected canvas, and the resulting 3D point lying on the canvas is stored as the continuation of the corresponding 3D stroke polyline vertices.}
    \label{fig:CANVASES_RAY_TRACING}
\end{figure}

The interface is composed of a ground plane within its environment and offers a choice of various geometric primitives as preset built-in canvases, including a plane, cube, sphere, and cylinder. The designer can select a canvas and arbitrarily transform (translate, rotate, and scale) it throughout the scene while having control over the camera viewpoint's position and rotation. With such controls, once the camera and canvas are in the designer's desired transformation, the designer can lock both and start drawing a stroke on the canvas from the respective locked viewpoint. In this setting, as the designer starts drawing a stroke on the tablet's surface, the ray originating from the camera viewpoint is intersected with the canvas's triangular mesh, and the resulting 3D point is stored as the continuation of the stroke polyline vertices, forming a 3D stroke (Figure \ref{fig:CANVASES_RAY_TRACING}). Depending on the chosen brush type, the resultant 3D stroke is either rendered in the form of triangle strips (ruled surface strips) or square sweeps of user-specified width (ranging continuously from $0.01$ to $1.0$) centered around the 3D stroke polyline vertices. It is noteworthy that the designer is not limited to use just one single canvas; the overall sketch can incorporate combinations of strokes with varying types of parent canvases drawn from varying viewpoints, enabling the design of complex architectural objects. To aid the designer in terms of precision and scale, they are provided with two canvas visualization methods interface: \textit{grid visualization}, which renders the drawing canvas with a grid of major (white) and minor (black) lines, where major lines are spaced 1 m apart and minor lines are spaced 10 cm apart; and \textit{intersection visualization}, which renders green lines indicating where the current active canvas intersects with previously sketched strokes.

Throughout the drawing process, along with the positions of the strokes' 3D polyline vertices and their corresponding timestamps, additional metadata comprising a set of geometry and stylus-related properties is actively recorded with the aim of facilitating data-driven sketch analysis. While some of these properties are recorded per each stroke, some others are recorded per each stroke polyline vertex. Properties that are attributed to each stroke include \textit{inkColour}, \textit{inkWidth}, \textit{cameraViewPosition}, \textit{cameraViewRotation}, \textit{canvasID}, and \textit{canvasTransform}. As their names imply, they represent the brush color and width chosen by the designer, the camera position, the camera rotation when drawing the stroke, the ID, and the transformation matrix of the stroke's parent canvas, respectively. Additionally, for every vertex point on the 3D stroke polyline, several properties are individually recorded, including \textit{normal}, \textit{tilt}, \textit{twist}, and \textit{pressure}. These represent the normal vector of the canvas at each point, the stylus's tilt in the $x$ and $y$ directions, the twist of the stylus, and the perpendicular force applied to the tablet's surface at each polyline vertex point of the drawn stroke, respectively.

\section{Input Drawing Characteristics}
\label{sec:INPUT_DRAWING_CHARACTERISTICS}

Our study of architectural sketching practices \cite{ching2019design}, together with close observation and analysis of sketches created by two modeling experts in our research group as well as those created by participants in our user study (Section \ref{sec:USER_STUDY}) point to several common core characteristics inherent in architectural design sketches:
\paragraph*{Distinct Types.} The designer-drawn strokes are mainly of two kinds, one depicting the form and structure, and the other depicting the tone and texture \cite{ching2019design}. For form and structure, designers often draw a set of strokes to outline the boundaries and edges of their intended architectural object, which we refer to as \textit{Shape} strokes in this paper. While these strokes are essential to communicate the design idea, the surfaces of forms and shapes cannot be fully described by such strokes alone. To this end, designers draw another set of tonal strokes to colorize, fill in, and mark the enclosed areas by the boundaries, forming the faces of the intended geometry. There are several basic techniques for creating tonal strokes, such as hatching, scribbling, and stippling; due to the great flexibility scribbling offers compared to others, we primarily focus on scribbling in this paper, referring to this particular type of strokes as \textit{Scribble} strokes (Figure \ref{fig:CUBE_INSPIRATION}). In this context, Shape strokes are usually delineated elaborately with more precision and intent in a single direction. On the other hand, Scribble strokes are often drawn as a network of random, multi-directional lines.

\paragraph*{Impreciseness.} When creating sketches, designers frequently draw multiple, tightly clustered, over-sketched strokes to depict their intended geometry. Also, designer-drawn strokes at junctions are often imprecise, with strokes intended to intersect ending up either short of doing so or over-shooting.

\begin{figure}[]
    \centering
    \includegraphics[width=0.750\linewidth]{FIGURES/SHAPE_SCRIBBLE_2D.png}
    \caption{An example of a conventional 2D drawing of an architectural object portraying two types of strokes as the design: strokes outlining the boundaries and those marking the enclosed areas. Images' source: iStock. Credit: SireAnko. Licenses purchased by the first author.}
    \label{fig:CUBE_INSPIRATION}
\end{figure}


\section{Strokes2Surface: The Geometry Reconstruction Pipeline}
\label{sec:STROKES2SURFACE_THE_GEOMETRY_RECONSTRUCTION_PIPELINE}

The described characteristics of sketches are analogous to how curve networks are formed and surfaced, thus motivating the idea of curve network recovery from architectural design sketches. By first recognizing the types of individual strokes in the sketch, potential rough and coarse Shape strokes can then be parsed into groups and consolidated. Hence, treating them as curve segments in the curve network facilitates curve network recovery once their pairwise connectivities are fixed. Similarly, Scribble strokes, when parsed into clusters, can correspond to cycles in the recovered curve network, which can subsequently be leveraged for surfacing the recovered network. In the following subsections we describe the individual steps in the curve reconstruction pipeline.


\subsection{Stroke Type Recognition: Shape Versus Scribble}
\label{sub_sec:STROKE_TYPE_RECOGNITION}

We observed subtle nuances in a designer's approach toward sketching boundaries and edges (Shape strokes) compared to the enclosed areas and faces (Scribble strokes), revealing a distinct pattern. Specifically, designers seemed to draw Shape strokes more deliberately with a slower pace, often as short straight lines, in contrast to Scribble strokes, which were often drawn loosely with a higher speed, featuring longer lines with numerous turning points. These characteristics suggest the possibility of using a stroke classifier for Shape versus Scribble strokes given the right input features. To this end, first, given the metadata recorded by the drawing interface during sketching per stroke and per stroke polyline vertex (Section \ref{sec:THE_4D_DRAWING_INTERFACE}), we employ this recorded metadata either directly as input features, or leverage them to extract a set of hand-engineered features (Section \ref{sub_sub_sec:FEATURE_EXTRACTION}). Second, we individually and independently evaluate the relevance of each feature with respect to its significance for predicting the binary target variables in our curated dataset: Shape ($1$) versus Scribble ($0$) (Section \ref{sub_sub_sec:FEATURE_SIGNIFICANCE_TESTING}). Ultimately, we perform ablation studies over four subsets of features and two candidate classifiers and select the best model and subset of features based on the reported metrics (Section \ref{sub_sub_sec:ABLATION STUDIES}).


\subsubsection{Feature Extraction}
\label{sub_sub_sec:FEATURE_EXTRACTION}

We extract a total of 11 features for each stroke in a sketch. The computation of some features relies solely on geometry-related properties of strokes, some depend solely on stylus-related properties, and others are a combination of both. In Table \ref{tab:FEATURE_EXTRACTION}, we outline the definition of each feature and explain the rationale behind their extraction.

\begin{table*}[]
    \centering
    \begin{tabular}{ m{1.95cm} | m{7.45cm} | m{7.00cm} }
        \multicolumn{1}{c|}{Feature} & \multicolumn{1}{c|}{Definition} & \multicolumn{1}{c}{Reason} \\
        \hline
        \hline
        \textit{AvgPressure} & Mean pressure applied when drawing the stroke computed using interface-recorded "pressure" property per stroke polyline vertices & Observed deliberation in drawing Shape strokes compared to Scribbles. \\
        \hline
        \textit{AvgSpeed} & Mean speed used to draw the stroke, calculated by computing the speed at each vertex using its coordinates and timestamp, and those of its subsequent vertex & Motiviated similarly as \textit{AvgPressure}. \\
        \hline
        \textit{AvgTilt} & The mean stylus tilt when drawing the stroke, computed using interface-recorded "tilt" property & Designers may slightly tilt the stylus when mark the enclosing areas, a gesture potentially less prevalent in Shape strokes. \\
        \hline
        \textit{ColorShift} & The mean $L_{2}$ distance between the InkColor of stroke $S_{i}$ and the InkColors of its each \textit{neighboring} stroke $S_{j}$ that is within a distance of $2.5 \times \frac{w_{i} + w_{j}}{2}$ & Shape strokes are often drawn with a different color than neighboring Scribbles strokes. \\
        \hline
        \textit{Density} & Number of stroke polyline vertices, once simplified using the Ramer-Douglas-Peucker (RDP) algorithm \cite{douglas1973algorithms}, divided by the original count of vertices ($\epsilon = 0.5 \times \text{InkWidth}$) & Scribble strokes require a great count of vertices after simplification due to their complexity and numerous turning points. \\
        \hline
        \textit{Dist} & The geodesic distance between the two endpoints of the stroke on its parent canvas & Scribble strokes often have endpoints close to each other, due to their due to their forward/backward movements. Hence, often resulting in lower Dist values. \\
        \hline
        \textit{Duration} & The time taken to draw the stroke, computed using the timestamps of the first and last points & Potentially higher in Scribble strokes to to their longer length. \\
        \hline
        \textit{Length} & The cumulative arc length of the stroke & Scribble strokes are generally observed to possess longer length. \\
        \hline
        \textit{Order} & Normalized stroke order in the sketch & designers are often observed outlining Shape strokes first before transitioning to Scribble strokes. \\
        \hline
        \textit{PrimSegCount} & Number of primary segmentation points of a stroke computed as described in \cite{fiorentino20033d} (dependent on \textit{AvgSpeed}) & Exhibiting higher values in Scribble strokes due to their increased count of segments, angles, and motion. \\
        \hline
        \textit{Straightness} & The \textit{Dist} feature divided by $\textit{Length}$ & Measuring stroke straightness, potentially resulting in higher values in Shape strokes. \\
        \hline
    \end{tabular}
    \caption{Extracted features, their corresponding definitions, and the reasons behind their inclusion.}
    \label{tab:FEATURE_EXTRACTION}
\end{table*}

\begin{figure*}[!h]
    \centering
    \includegraphics[width=1.00\linewidth]{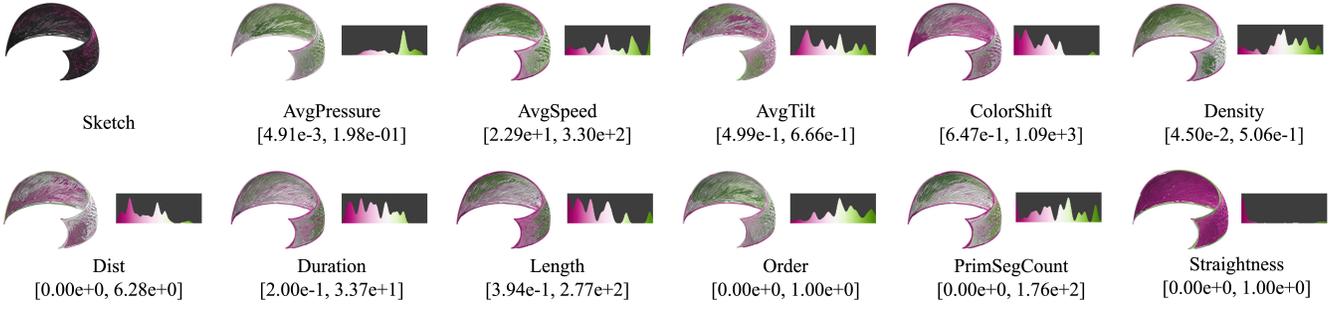}
    \caption{The top-left image represents the original sketch of a curved wall, created using the 4D drawing interface for theater stage design purposes, and the rest are the visualizations of raw, scalar-valued features of the sketch strokes that are color-coded from pink to green using a colormap that maps feature values into distinct colors according to their magnitude. The gradient ranging from pink to green denotes the increase in feature values, with pink representing lower values and green representing higher values. Below each image, $\min$ and $\max$ of each feature value are noted. Input sketch: Ingrid Erb.}
    \label{fig:FEATURES_TRENDS}
\end{figure*}

To gain additional visual insight into how the values of these features differ between Shape and Scribble strokes, Figure \ref{fig:FEATURES_TRENDS} displays their trends in a sample sketch. 

\subsubsection{Stroke Classifier}
\label{sub_sub_sec:STROKE_CLASSIFIER}

\begin{figure}[]
    \centering
    \includegraphics[width=1.00\linewidth]{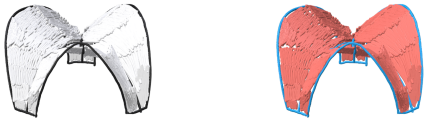}
    \caption{Left: the original sketch if a pavilion with three legs. Right: the classifier's predictions on strokes with Shape Strokes colored in blue and Scribble strokes in red.}
    \label{fig:STROKE_TYPE_RECOGNITION}
\end{figure}

The trends in stroke feature values on boundaries compared to enclosed areas suggest their potential use in a classification task. After statistical analysis (Section \ref{sub_sub_sec:FEATURE_SIGNIFICANCE_TESTING}) and excluding AvgPressure and AvgTilt due to their insignificant relevance to the binary target variables, we proceed with our best meta-estimator model ($\text{RF}_{\mathcal{GEO}\vee\mathcal{STY}}$) after performing ablation studies (Section \ref{sub_sub_sec:ABLATION STUDIES}). We use this random forest classifier for inference, with its predictions on a sample unseen sketch shown in Figure \ref{fig:STROKE_TYPE_RECOGNITION}.


\subsection{Curve Network Formation}
\label{sub_sec:CURVE_NETWORK_FORMATION}

Upon the removal of Scribble strokes from the classifier's predictions on the input sketch, we are left with Shape strokes delineating the boundaries and edges of the desired geometry. Typically, they are expressed by multiple tightly clustered strokes that may be partly or fully over-sketched or sketched as a continuation of one another by the designer. Thus, recovering a curve network from the remaining strokes requires several steps; it is necessary to further cluster Shape strokes into separate groups, beautify and consolidate them, and recover the pairwise connectivities among consolidated segments.


\subsubsection{Clustering Shape Strokes}
\label{sub_sub_sec:CLUSTERING_SHAPE_STROKES} 

In our setting, Shape strokes depicting the same edge cannot always be consolidated using existing techniques in 2D \cite{liu2018strokeaggregator, van2021strokestrip}. This is mainly due to two reasons: firstly, as pointed out in \cite{rosales2019surfacebrush}, 2D sketched stroke vertices have unique nearest left/right neighbors along the stroke's orthogonal. This property no longer holds in 3D, making the determination of the best pairwise vertex matches a lot more challenging. Secondly, and more specific to our problem, it is observed that even Shape strokes depicting the same edge do not always have the same parent canvas geometry to use 2D consolidation methods on the canvas's surface. As an example, assume the designer drawing a cube using solely a plane canvas; a commonly observed scenario is that when drawing two adjacent faces meeting at a shared edge, sometimes the designer depicts the edge by partially over-sketching strokes lying on different planes, making consolidation on the canvases' surfaces inapplicable. These two limitations necessitate a sketch consolidation method tailored to 3D sketches. 

In order to detect stroke groups for consolidation, we compute a pairwise similarity score between Shape strokes. However, prior to that, strokes must go through a pre-processing process for two reasons: first, there is often redundancy and noise within the recorded stroke polyline vertices because of the varying drawing speed and the nature of the sampling strategy in the sketching software. Second, stylus slippage on the tablet surface often results in unintended hooks at the beginning and end of the strokes. To this end, using the method described in \cite{fiorentino20033d}, we filter the Shape stroke polyline vertices to eliminate possible redundancy and noise, and then approximate the stroke by fitting a cubic B-Spline curve to each stroke. 

After pre-processing, to be able to compute the similarity scores between each pair of Shape strokes $S_{i}$ and $S_{j}$, where $1 \leq i, j \leq M$ and $M$ is the total number of strokes classified as Shape, we switch to a point-based representation of the obtained curves by sampling a fixed number of points inversely proportional to the curves' lengths, denoted as $P_{i}$ and $P_{j}$. After turning the polyline vertices of each stroke into a K-Dimensional Tree (KDTree), matching sequences between each pair of strokes are computed. We define a matching sequence between strokes $S_{i}$ and $S_{j}$ as a pair of subsets of ordered points $M_{k} = (Q_{k, i}, Q_{k, j})$ where $Q_{k, i} \subseteq P_{i}$ and $Q_{k, j} \subseteq P_{j}$, such that for each point $q_{k, i} \in Q_{k, i}$, there exists a point $q_{k, j} \in Q_{k, j}$ such that the Euclidean distance between $q_{k, i}$ and $q_{k, j}$ is at most $1.5 \times \text{min} (w_{i}, w_{j})$. Here, $w_{i}$ and $w_{j}$ denote the stroke widths of $S_{i}$ and $S_{j}$, respectively. 

Given the definition above, for every two curves that satisfy the above criteria, we obtain a set of matching sequences $\mathcal{M}_{i, j} = \{M_{1}, \cdots, M_{\mid \mathcal{M} \mid} \}$, where each $M_{i}$ is a tuple containing a subset of ordered points in $P_{i}$ and $P_{j}$. Thereafter, for each matching sequence $M_{k}$, where $1 \leq k \leq \mid \mathcal{M} \mid$, let $\bar{t}_{k, i}$ and $\bar{t}_{k,j}$ be the vectors denoting the average tangent directions along $Q_{k, i}$ and $Q_{k, j}$, respectively. Pairwise computation of the dot products between the tangents for each pair matching sequence in $M_{k}$ leads to the set $\{\bar{t}_{1, i} \cdot \bar{t}_{1, j}, \cdots, \bar{t}_{\mid \mathcal{M} \mid, i} \cdot \bar{t}_{\mid \mathcal{M} \mid, j}  \}$. Taking the average of all such dot products over all matching sequences $M_{k}$ between the two strokes will lead to a number between $0$ and $1$, representing the similarity score of the two strokes:
\begin{equation}
    \textnormal{Score}_{\textnormal{\tiny Shape}}(S_{i}, S_{j}) = \frac{1}{\mid \mathcal{M} \mid} \sum_{k=1}^{\mid \mathcal{M} \mid} \bar{t}_{k, i} \cdot \bar{t}_{k, j}.
\end{equation}
If there exists no matching sequence between a pair of strokes, we set the score to $0$. It is crucial to highlight that the coefficient $1.5$ is chosen empirically and deliberately set to a number higher than $1$ to account for the situations where the two strokes are meant to overlap or continue each other but slightly stop short of doing so. 

\begin{figure}[]
    \centering
    \includegraphics[width=1.\linewidth]{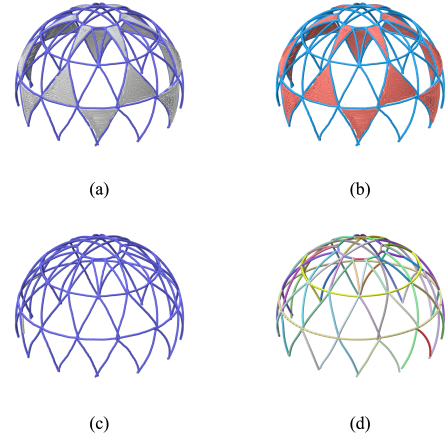}
    \caption{(a) A sample sketch of a dome-like architectural object created by participant P01. (b) The outcome of the classification process, where strokes predicted as Scribble are highlighted in red, and those identified as Shape are marked in blue. (c) The sketch after removal of the Scribble strokes, leaving only the Shape strokes in their original color. (d) The result of clustering the Shape strokes into distinctive groups, each representing a boundary or edge, with each group displayed in a unique color for clarity.}
    \label{fig:CLUSTERING_SHAPE_STROKES_01}
\end{figure}

The pairwise similarity scores computed above are used in conjunction with the DBSCAN \cite{ester1996density} clustering algorithm with the $\epsilon$ value determined dynamically based on its $k$-distance plot with $k=1$ (refer to Supplementary Material for more details). Figures \ref{fig:CLUSTERING_SHAPE_STROKES_01}a and \ref{fig:CLUSTERING_SHAPE_STROKES_01}d show a sample input sketch and the resultant clusters of the Shape strokes following the approach above.

\begin{figure}[]
    \centering
    \includegraphics[width=.95\linewidth]{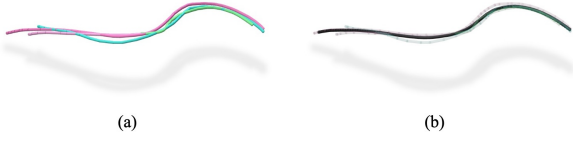}
    \caption{Consolidation results. (a) Shows multiple over-sketched curves, each rendered using a random color assigned to it. (b) The consolidated curve overlaid on the original strokes with reduced transparency.}
    \label{fig:CLUSTERING_SHAPE_STROKES_02}
\end{figure}

\paragraph*{Consolidation.} Once the clusters are identified, they must be further consolidated into a single curve. Our clustering method computes the pairwise scores between shape strokes based on their local similarities. However, at times, the strokes in an identified group may feature bifurcating branches, representing two or more edges that form a Y-junction, thereby requiring the formation of two or more curves to approximate a cluster instead of one. Such cases may be difficult to detect at a local level, necessitating global post-processing of each group. Inspired by similar works in 2D \cite{orbay2011beautification, liu2018strokeaggregator, van2021strokestrip}, as of the post-processing, we detect such cases and further divide the clusters exhibiting Y-junctions into sub-clusters exhibiting no bifurcation points. To identify these cases, we initially thin each cluster's point cloud -- the set of points of the strokes falling into the cluster -- using the method of \cite{lee2000curve}, employing the modified Moving-Least-Squares (MLS) fitting algorithm extended to 3D. Thinning the point cloud requires a parameter $H$ that denotes the initial thickness of the point cloud. We set this parameter to the mean of the brush widths of the cluster's constituent strokes. Once a sufficiently thin point cloud is obtained, we compute the Euclidean Minimum Spanning Tree (EMST) $T$ of the thinned point cloud to check for bifurcating points. Similar to the 2D case \cite{orbay2011beautification}, the candidate bifurcating points are those sets of points in the tree that have at least three neighboring points and that the sub-graphs and branches created by removing such points are significant enough in terms of their size and length. Whenever $\frac{T_{\min}}{T_{\max}} \geq 0.05$, the point is marked as a candidate bifurcating point, where $T_{\min}$ and $T_{\max}$ are the minimum and maximum cumulative subgraph lengths formed by removing the candidate point. By iterating over thinned cluster points, identifying bifurcating points, and subsequently splitting the cluster, a series of sub-clusters are obtained. Each sub-cluster represents a single Y-junction-free boundary and edge of the intended geometry. Finally, before recovering the intended topology to form a curve network, the orderings of the points along each (sub-) cluster must be known for further curve approximation. To this end, we use the EMST of each sub-cluster and traverse the shortest path between the two ends of the tree and order the points. Finally, the ordered points are approximated with a cubic B-spline curve with four control points (Figure \ref{fig:CLUSTERING_SHAPE_STROKES_02}).

\subsubsection{Topology Recovery}
\label{sub_sub_sec:TOPOLOGY_RECOVERY}

The obtained consolidated curves representing the boundaries and edges are still disconnected and should be further processed for the creation of a well-connected curve network. To this end, for each curve $C_{i}$ we compute its shortest distance to every other curve $C_{j}$, where $1 \leq i, j \leq N $, $i \neq j$, and $N$ is the total number of consolidated curves obtained from the preceding processes:
\begin{align}
    d_{i, j} &= \min \parallel p_{i, u} - p_{j, v} \parallel \nonumber \\
    &\begin{aligned}
        &\text{s.t.} \quad 1 \leq u \leq \text{Num}_{i}, \\
        &\hspace{0.4cm} \quad 1 \leq v \leq \text{Num}_{j}
    \end{aligned}
\end{align}
where $p_{i, u}$ and $p_{j, v}$ represent sampled points on the curves' parameter space with $\text{Num}_{i}$ and $\text{Num}_{j}$ being the total number of such points on the curves $C_{i}$ and $C_{j}$, respectively. If the $d_{i, j}$ is less than $1.5 \times \text{min} (w_{i}, w_{j})$, we deem the curves as requiring a connection. We then check if the point 
\begin{equation}
    p_{i, u}^{*} = \underset{i, u}{\arg \min} \hspace{0.125cm} d_{i, j}
\end{equation}
is sufficiently close to one of the endpoints of the curve $C_{i}$, in which case we constrain the corresponding endpoint; otherwise, we split the curve $C_{i}$ at $p_{i, u}^{*}$ and constrain the end-points of the resultant curves $C_{i}^{1}$ and $C_{i}^{2}$ to be connected to the point $p_{i, u}^{*}$. By repeating this process for each curve, we obtain a new set of curves that either do not require a connection or are constrained on their endpoints to be connected to a target connecting point. Given a subset of curves $\hat{C}_{1}, \cdots, \hat{C}_{N}$ with their corresponding endpoints $\hat{p}_{1}, \cdots, \hat{p}_{N}$ constrained to be connected to a target connecting point, we first compute the tangent vectors $\hat{t}_{1}, \cdots, \hat{t}_{N}$ for each endpoint on each curve. We then determine their connecting point by minimizing the following function using the Broyden-Fletcher-Goldfarb-Shanno (BFGS) method:
\begin{equation}
    \sum_{i = 1}^{N} \parallel ({\hat{p} - \hat{p}_{i}}) \times \hat{t}_{i} \parallel
\end{equation}
Once the optimum point $\hat{p}^{*}$ is determined, each of the curves $\hat{C}_{i}$ is extended from their $\hat{p}_{i}$ end to the point $\hat{p}^{*}$, and once again, we approximate the new updated set of points on each curve $\hat{C}_{i}$ with a cubic B-Spline. It should be noted that when computing the shortest distance between two curves, the coefficient $1.50$ was empirically selected, as this particular value has demonstrated the best performance across the set of sketches experimented with. 

\begin{figure}[]
  \centering
  \includegraphics[width=0.825\linewidth]{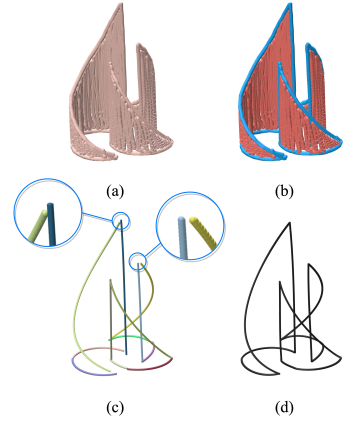}
  \caption{(a) An exemplar sketch created by the user study participant P02, utilizing the cylinder canvas. (b) The results obtained from the classification process. (c) The consolidated Shape sub-clusters derived following the post-processing stage; note the insets highlighting the disconnections amid the curves. (d) Depicts the output of the topology recovery process, resulting in a well-connected curve network.}
  \label{fig:TOPOLOGY_RECOVERY}
\end{figure}

Following the process above, we obtain a set of curves that are well-connected, thereby forming a well-connected curve network(s) for the sketch (Figure \ref{fig:TOPOLOGY_RECOVERY}). Next, we need to infer which cycles of the curve network should bound a surface patch and which should not. To this end, we rely on the Scribble strokes groups, as described in the next section.

\subsection{Surfacing Curve Network}
\label{sub_sec:SURFACING_CURVE_NETWORK}

Finding cycles in a curve network is generally a complex and ambiguous problem. Several methods have been developed to automatically find an acceptable set of cycles that bound patches, often involving global optimization over the entire curve network \cite{zhuang2013general, abbasinejad2012surface, stanko2016smooth}. However, in our problem, we can leverage the presence of Scribble strokes to more accurately discover the designer-intended cycles in the recovered curve network. To achieve this, we can cluster the Scribble strokes into distinct groups, each representing a single face of the intended geometry of the architectural object. Then, we can apply the existing algorithms locally to the neighboring boundary curves of each group — the segments of the curve network that fall within the bounding box of that group — rather than running them globally on the complete curve network.


\subsubsection{Clustering Scribble Strokes}
\label{sub_sub_sec:CLUSTERING_SCRIBBLE_STROKES}

We cluster Scribble strokes using a similar strategy to the one employed for the Shape strokes (Section \ref{sub_sub_sec:CLUSTERING_SHAPE_STROKES}). The designers are mainly observed to draw single or multiple overlapping Scribble strokes on the same canvas with the same transformation matrix to fill in an enclosing area by boundary curves. However, exceptions occasionally occur when the designer depicts a single face with multiple canvases of the same type or of different types but with distinct transformation matrices -- different positions, scales, and rotations. Thus, a general and robust clustering approach for Scribble strokes should account for such scenarios. Given this, for every two Scribble strokes $S_{i}$ and $S_{j}$, we first compute the total number of their polyline points that are at most in distance $1.5 \times \text{min} (w_{i}, w_{j})$ of each other. Assuming that $N_{i, j}$ shows the cardinality of all such pairs, their corresponding similarity score is computed as follows:
\begin{equation}
    \begin{split}
        \textnormal{Score}_{\textnormal{\tiny Scribble}}(S_{i}, S_{j}) = & \min (C \times 12.5 \times \frac{N_{i,j}}{N_{i} + N_{j}}, 1) \\
        & - 25 \times \min_{1 \leq k \leq N} N_{i, j, k}
    \end{split}
    \label{eq:SCRIBBLE_CLUSTERING_01}
\end{equation}
where $\text{N}_{i}$ and $\text{N}_{j}$ denote the number of points on Scribble strokes $S_{i}$ and $S_{j}$. $C$ is the coefficient taking into account the canvas types and their corresponding transformation at the time of drawing the two Scribble strokes. Assuming that $CP_{i}$ and $CP_{j}$ denote the coordinates of the center of the two parent canvases $C_{i}$ and $C_{j}$ of the two strokes $S_{i}$ and $S_{j}$, C is computed as follows
\begin{equation}
    C = \begin{cases}
        100, & \text{Type}(C_{i}, C_{j}) = \text{Plane} \land \frac{1}{\parallel CP_{i} - CP_{j} \parallel} \leq 0.375\\
        100, & \text{Type}(C_{i}, C_{j}) = \text{Sphere} \land \frac{1}{\parallel CP_{i} - CP_{j} \parallel} \leq 1.750\\
        \frac{1}{12.5}, & \text{otherwise}
    \end{cases}
    \label{eq:SCRIBBLE_CLUSTERING_02}
\end{equation}
Moreover, the second term in the equation is to account for the cases where a shape stroke is going through the middle of the two Scribble strokes $S_{i}$ and $S_{j}$. In such cases, it is likely that the Scribble strokes must belong to two different faces of the intended geometry as they are being separated by a Shape stroke. To this end, we simply compute the fraction of points of stroke $N_{i, j}$ that are overlapping with each stroke classified as Shape earlier and compute the numbers $N_{i, j, k}$ for $1 \leq k \leq N$, where $N$ is the total number of strokes classified as Scribble. 

\begin{figure}[]
  \centering
  \includegraphics[width=0.625\linewidth]{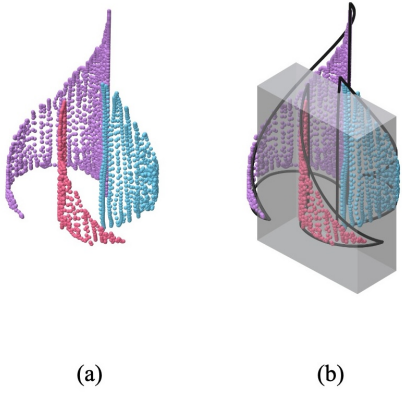}
  \caption{(a) The Scribble strokes shown in Figure \ref{fig:TOPOLOGY_RECOVERY}b parsed into clusters, each representing a single face of the geometry. (b) The axis-aligned bounding box for one of the Scribble clusters of the sketch. Only the subset of curve segments forming the boundaries of the cluster colored in pink fall into the bounding box area and are used for cycle discovery.}
  \label{fig:SURFACING_CURVE_NETWORK}
\end{figure}

Similar to Shape clustering, using the pairwise scores defined above, we merge clusters using the DBSCAN algorithm with the $\epsilon$ parameter determined dynamically for each sketch (refer to Supplementary Material for more detail). Figure \ref{fig:SURFACING_CURVE_NETWORK}a shows the resultant Scribble clusters computed following the above method. 

\subsubsection{Cycle Discovery and Geometry Generation} 
\label{sub_sub_sec:CYCLE DISCOVERY_AND_GEOMETRY_GENERATION}

Parsed Scribble clusters explicitly represent the faces of the intended architectural object and can be exploited to identify the cycles in the curve network that must bound patches. To this end, we compute the axis-aligned bounding box for the points of the strokes forming a Scribble cluster and then scale the bounding box slightly with a factor of $1.50$. We then take the subset of curves falling into the bounding box and use it as input to the method presented in \cite{zhuang2013general} to find it corresponding formed cycles cycles (Figure \ref{fig:SURFACING_CURVE_NETWORK}). Next, we triangulate the closed 3D area formed by each cycle's boundary curves using the method described in \cite{zou2013algorithm}. Repeating this process for every Scribble cluster leads to all the cycles in the curve network that are must-bound patches. Last, sometimes, some unwanted curves fall into the bounding box and lead to extra potentially wrong patches. To address this, after surfacing all the identified cycles, we discard any bounding patch where no cluster group can be projected onto the surface using its normal vector's direction and distance thresholding. We note that the scaling factor is intentionally chosen to a number greater than $1$ to ensure boundary curves fall within the bounding box, even if Scribbles are not densely sketched.


\section{User Study}
\label{sec:USER_STUDY}

The purpose of our user study was three-fold. First, we aimed to confirm further our studies and observations (Section \ref{sec:INPUT_DRAWING_CHARACTERISTICS}) regarding the characteristics of architectural design sketches on which the pipeline is based and developed. Second, to collect a dataset of such sketches to train our model's binary classifier. And third, to validate and assess the usability of the pipeline in architectural design use cases. To this end, we conducted a workshop with 10 participants (P01-P10), where the majority of them were university students of architecture or civil engineering programs and were experienced with at least one commercial 3D design modeling software (refer to Supplementary Materials for the individual participants' design backgrounds).

Initially, participants were given a 15-minute presentation introducing the research project and providing an overview of the drawing interface and the Strokes2Surface geometry reconstruction pipeline. This was followed by a 30-minute tutorial that familiarized the participants with the various features of MR.Sketch in detail and the framework of Strokes2Surface. Throughout the tutorial, each student was guided by the instructor to sketch a \textit{3D cube} for a hands-on experience with the MR.Sketch (the given instruction can be found in Supplementary Materials). After this initial getting-started session, once the participants felt comfortable with the sketching interface, they were tasked with drawing two architectural objects within a 1-hour time frame. First, they were tasked to draw from \textit{observation}, where each participant had to select an object from a pre-curated list of actual built structures in the real world (refer to Supplementary Material for the links to the selected structures). For each structure, they were provided with reference images from various views, and these images remained visible on a large screen throughout the session to offer clear drawing targets. Second, they were tasked to draw from \textit{imagination}, where the users could sketch a structure by following their own ideation process. Depending on the complexity and scale with which the designer sketched the objects, it took the participants varying amounts of time to complete the tasks. Nevertheless, all the participants completed both tasks within the allotted time frame. Moreover, five of the participants were keen and chose to stay for about an additional hour to draw extra objects.


\subsection{Preliminary Examination}
\label{sub_sec:PRELIMINARY_EXAMINATION}

In line with our first aim, our examination of each sketch's constituent strokes revealed that participants created sketches analogous to conventional 2D architectural sketching; they often drew boundaries, whilst they often filled in the enclosing areas to block out strokes seen in the background. We were also excited to observe slight variations in sketching styles among participants. For example, we observed slight variations in the density of their sketches, with some participants sketching objects more densely and others more sparsely. Furthermore, some participants drew a single long stroke to depict multiple connecting boundary curves, in contrast to others who favored multiple shorter strokes in similar situations. Users also varied in their use of color to delineate boundaries; some employed multiple colors, while others preferred using a single color across the sketch (Figure \ref{fig:TOPOLOGY_RECOVERY}a). The extent of over-sketching varied among participants as well; some tended to emphasize on previously drawn boundary strokes with over-sketching, whereas others did not over-sketch as much. Overall, these observations were crucial to the development of Strokes2Surface to ensure robustness regarding a wider variety of sketches and styles and also provided us valuable insights for future work.


\subsection{Dataset}
\label{sub_sec:DATASET}

Following our second aim, we further manually labeled strokes of each sketch with one of three ground-truth values: $0$ for Scribble, $1$ for Shape, and $-1$ as noise for those that were unclear or created using techniques other than scribbling. These sketches created by participants throughout the study and eight additional sketches made by two expert architectural designers from our research group were curated as a dataset of freehand 4D architectural sketches. Overall, it contains 47 architectural labeled sketch drawings encompassing 4284 4D strokes coupled with their additional metadata recorded by the drawing interface (Section \ref{sec:THE_4D_DRAWING_INTERFACE}). Table \ref{tab:DATASET_STATISTICS} presents the dataset statistics, categorized by the mode of drawing, i.e., 3D cube, drawings from observation, or the imagination.

\begin{table}[]
    \centering
    \begin{tabular}{c | c | c}
        & Sketches & (Shape / Scribble / Noise)\\
        \hline
        \hline
        3D Cube & 10 & 635 (375 / 224 / 36)\\
        \hline
        Observation & 10 & 1973 (1024 / 906 / 43)\\
        \hline
        Imagination & 27 & 1676 (1029 / 623 / 24)\\
        \hline
        Sum & 47 & 4284 (2428 / 1753 / 103)\\
        \hline
    \end{tabular}
    \caption{Statistics of the collected dataset of 4D architectural design sketches.}
    \label{tab:DATASET_STATISTICS}
\end{table}


\subsection{User Feedback}
\label{sub_sec:USERS_FEEDBACK}

Finally, once we trained and tested the classifier using the labeled data, we provided each participant a customized link to a web-based interface, allowing them to visualize from arbitrary viewpoints the color-coded step-wise outputs of the pipeline as it ran on their individual sketches. Along with the link, they received a post-study questionnaire designed to quantitatively and qualitatively probe their assessment of the interface (MR.Sketch) and the pipeline's interpretations Strokes2Surface).

The data from our questionnaires are presented in Figures \ref{fig:QUESTIONS_MRSKETCH} and \ref{fig:QUESTIONS_STROKES2SURFACE}. Aggregating responses across all questions relating to MR.Sketch (Q1-Q5) and those relating to Strokes2Surface (Q6-Q9), followed by one-sample non-parametric permutation test against the theoretical median response of "Neutral", indicates a significant effect in favor of MR.Sketch ($p\text{-value} \approx 0.029 < 0.05$) and Strokes2Surface ($p\text{-value} \approx 0.005 < 0.05$).

In the open-ended feedback sections, participants reflected on their experience in transferring their 2D drawing skills to 3D sketching with MR.Sketch: \textcolor{gray}{"In contrast to 2D sketching, which is single-perspective, here I had the freedom over the entirety of my sketch, which made me more confident to design (P01)"}, \textcolor{gray}{"... MR.Sketch allowed me to sketch quickly and roughly like I am used to when making concept sketches in 2D (P09)"}. In this regard, some also pointed out how the tutorial session helped them to better understand the interface: \textcolor{gray}{"After 15-20 minutes I got used to it and was surprised by how easy it was to sketch 3D objects (P04)"}, and \textcolor{gray}{"In the beginning it was a bit difficult to get a grasp on it, but you quite easily get accustomed to it after tutorial (P07)"}. In response to a question asking participants for further suggestions or feedback to share with respect to MR.Sketch, they brought our attention to certain areas for improvement: \textcolor{gray}{"It would have been easier if I could copy part of my drawings, at least the boundaries (P01)"}, and two other participants commented on ease of use of canvas transformations: \textcolor{gray}{"I think this would boost the drawing experience if I could auto-snap the canvas to certain positions or pre-select an axis where it snaps to (P06)"}, and \textcolor{gray}{"being able to enter the exact angle I want to rotate the canvas, with just a few changes, I think it would be a very helpful tool to sketch in 3D (P03)"}. Alongside the raised points, users are currently restricted to built-in preset canvases and cannot create their own. Including this feature in future work could enhance the drawing experience, especially for freeform geometries. Likewise, saving and displaying previously used canvas positions would aid in more accurate sketching for adding strokes later on. 

Users were also asked to share their opinion on Strokes2Surface: \textcolor{gray}{"It would make it more convenient if Strokes2Surface could result in geometrically regular shapes (P03)"}. Although the pipeline reconstructs neat geometries, some properties such as symmetry are not enforced. This could be an intriguing direction for future research. Users elaborated further on how they see Strokes2Surface benefiting them: \textcolor{gray}{"... to explain designs to non-experts and customers I find it a very good system (P07)"}, \textcolor{gray}{"it is more efficient in externalizing architectural ideas than dealing with the complexity of CAD systems, especially when presenting ideas in the early stages of a design project (P09)"}, and a participant pinpointed another area which could be addressed in future work: \textcolor{gray}{"In the reconstructed geometry, if the system could recognize the type of building elements in sketch, it would really help with streamlining the design to geometry and further prepare for structural analysis (P08)"}.

\begin{figure*}
    \centering
    \begin{tikzpicture}
        \begin{axis}[
            width=13.00cm,
            height=12cm,
            bar width=8pt,
            y=25pt,
            symbolic y coords={
                {(Q5) I could easily use MR.Sketch to draw from the imagination.},
                {(Q4) I could easily use MR.Sketch to draw from observation.},
                {(Q3) I did not fill limited by the preset canvases.},
                {(Q2) It was easy to transform camera and canvases around the scene.},
                {(Q1) It was intuitive to use interface for 3D concept design.}
            },
            xbar stacked,
            xmin=0,
            xmax=100,
            xtick={0,20,...,100},
            xticklabels={0\%,20\%,40\%,60\%,80\%,100\%},
            ytick=data,
            yticklabel style={
                font=\small,
                inner sep=2pt,
                fill=white,
                text width=4.5cm,
                align=left,
                xshift=-10pt
            },
            legend style={
                at={(0.5,-0.15)},
                anchor=north,
                legend columns=-1,
                draw=none,
                font=\small,
                column sep=10pt,
                legend cell align=left
            },
            axis line style={draw opacity=0.1},
            xmajorgrids,
            grid style={
                color=gray,
                line width=0.25pt
            }]
            \addplot+[xbar, fill={rgb,1:red,1.00;green,0.27;blue,0.22}, draw=none] plot coordinates {
                (00.00,{(Q1) It was intuitive to use interface for 3D concept design.}) 
                (00.00,{(Q2) It was easy to transform camera and canvases around the scene.}) 
                (00.00,{(Q3) I did not fill limited by the preset canvases.}) 
                (00.00,{(Q4) I could easily use MR.Sketch to draw from observation.}) 
                (00.00,{(Q5) I could easily use MR.Sketch to draw from the imagination.})
            };
            \addlegendentry{Strongly disagree} 

            \addplot+[xbar, fill={rgb,1:red,1.00;green,0.62;blue,0.03}, draw=none] plot coordinates {
                (00.00,{(Q1) It was intuitive to use interface for 3D concept design.}) 
                (00.00,{(Q2) It was easy to transform camera and canvases around the scene.}) 
                (20.00,{(Q3) I did not fill limited by the preset canvases.}) 
                (10.00,{(Q4) I could easily use MR.Sketch to draw from observation.})
                (00.00,{(Q5) I could easily use MR.Sketch to draw from the imagination.})
            };
            \addlegendentry{Disagree} 

            \addplot+[xbar, fill={rgb,1:red,0.59;green,0.59;blue,0.61}, draw=none] plot coordinates {
                (10.00,{(Q1) It was intuitive to use interface for 3D concept design.}) 
                (30.00,{(Q2) It was easy to transform camera and canvases around the scene.}) 
                (30.00,{(Q3) I did not fill limited by the preset canvases.}) 
                (20.00,{(Q4) I could easily use MR.Sketch to draw from observation.})
                (10.00,{(Q5) I could easily use MR.Sketch to draw from the imagination.})
            };
            \addlegendentry{Neutral} 

            \addplot+[xbar, fill={rgb,1:red,0.35;green,0.78;blue,0.96}, draw=none] plot coordinates {
                (70.00,{(Q1) It was intuitive to use interface for 3D concept design.}) 
                (60.00,{(Q2) It was easy to transform camera and canvases around the scene.}) 
                (30.00,{(Q3) I did not fill limited by the preset canvases.}) 
                (50.00,{(Q4) I could easily use MR.Sketch to draw from observation.})
                (70.00,{(Q5) I could easily use MR.Sketch to draw from the imagination.})
            };
            \addlegendentry{Agree} 

            \addplot+[xbar, fill={rgb,1:red,0.03;green,0.51;blue,1.00}, draw=none] plot coordinates {
                (20.00,{(Q1) It was intuitive to use interface for 3D concept design.}) 
                (10.00,{(Q2) It was easy to transform camera and canvases around the scene.}) 
                (20.00,{(Q3) I did not fill limited by the preset canvases.}) 
                (20.00,{(Q4) I could easily use MR.Sketch to draw from observation.})
                (20.00,{(Q5) I could easily use MR.Sketch to draw from the imagination.})
            };
            \addlegendentry{Strongly agree} 
        \end{axis}
    \end{tikzpicture}
    
    \caption{Visualizations of the MR.Sketch questions. The questions are provided on the left. For each question there is a stacked bar where each bar has five responses from a 5-point Likert scale.}
    \label{fig:QUESTIONS_MRSKETCH}
\end{figure*}

\begin{figure*}
    \centering
    \begin{tikzpicture}
    \begin{axis}[
        width=13.00cm,
            height=12cm,
            bar width=8pt,
            y=25pt,
            symbolic y coords={
                {(Q9) Grouped Shape strokes are correctly connected.}, 
                {(Q8) Scribble strokes were correctly grouped.}, 
                {(Q7) Shape strokes were correctly grouped.}, 
                {(Q6) My sketched strokes are correctly interpreted/classified.}
            },
            xbar stacked,
            xmin=0,
            xmax=100,
            xtick={0,20,...,100},
            xticklabels={0\%,20\%,40\%,60\%,80\%,100\%},
            ytick=data,
            yticklabel style={
                font=\small,
                inner sep=2pt,
                fill=white,
                text width=4.5cm,
                align=left,
                xshift=-10pt
            },
            legend style={
                at={(0.5,-0.15)},
                anchor=north,
                legend columns=-1,
                draw=none,
                font=\small,
                column sep=10pt,
                legend cell align=left
            },
            axis line style={draw opacity=0.1},
            xmajorgrids,
            grid style={
                color=gray,
                line width=0.25pt
            }]
            \addplot+[xbar, fill={rgb,1:red,1.00;green,0.27;blue,0.22}, draw=none] plot coordinates {
                (00.00,{(Q6) My sketched strokes are correctly interpreted/classified.}) 
                (00.00,{(Q7) Shape strokes were correctly grouped.}) 
                (00.00,{(Q8) Scribble strokes were correctly grouped.}) 
                (00.00,{(Q9) Grouped Shape strokes are correctly connected.}) 
            };
            \addlegendentry{Strongly disagree} 
            
            \addplot+[xbar, fill={rgb,1:red,1.00;green,0.62;blue,0.03}, draw=none] plot coordinates {
                (00.00,{(Q6) My sketched strokes are correctly interpreted/classified.}) 
                (00.00,{(Q7) Shape strokes were correctly grouped.}) 
                (00.00,{(Q8) Scribble strokes were correctly grouped.}) 
                (00.00,{(Q9) Grouped Shape strokes are correctly connected.})
            };
            \addlegendentry{Disagree} 

            \addplot+[xbar, fill={rgb,1:red,0.59;green,0.59;blue,0.61}, draw=none] plot coordinates {
                (00.00,{(Q6) My sketched strokes are correctly interpreted/classified.}) 
                (00.00,{(Q7) Shape strokes were correctly grouped.}) 
                (00.00,{(Q8) Scribble strokes were correctly grouped.}) 
                (00.00,{(Q9) Grouped Shape strokes are correctly connected.})
            };
            \addlegendentry{Neutral} 

            \addplot+[xbar, fill={rgb,1:red,0.35;green,0.78;blue,0.96}, draw=none] plot coordinates {
                (20.00,{(Q6) My sketched strokes are correctly interpreted/classified.}) 
                (60.00,{(Q7) Shape strokes were correctly grouped.}) 
                (40.00,{(Q8) Scribble strokes were correctly grouped.}) 
                (40.00,{(Q9) Grouped Shape strokes are correctly connected.})
            };
            \addlegendentry{Agree} 

            \addplot+[xbar, fill={rgb,1:red,0.03;green,0.51;blue,1.00}, draw=none] plot coordinates {
                (80.00,{(Q6) My sketched strokes are correctly interpreted/classified.})
                (40.00,{(Q7) Shape strokes were correctly grouped.})
                (60.00,{(Q8) Scribble strokes were correctly grouped.})
                (60.00,{(Q9) Grouped Shape strokes are correctly connected.})
            };
            \addlegendentry{Strongly agree} 
        \end{axis}
    \end{tikzpicture}
    \caption{Visualizations of the Strokes2Surface questions structured similarly as Figure \ref{fig:QUESTIONS_MRSKETCH}.}
    \label{fig:QUESTIONS_STROKES2SURFACE}
\end{figure*}


\section{Results and Validation}
\label{sec:RESULTS_AND_VALIDATION}


\subsection{Training the Classifier}
\label{sub_sec:CLASSIFIER_TRAINING}

We quantify extracted features (Section \ref{sub_sub_sec:FEATURE_EXTRACTION}) using statistical analysis to ensure that only relevant features are retained for our classification. We also perform ablation tests, evaluating 8 different training configurations with 2 meta-estimators, each trained on 4 different subsets of the retained features. This is done to study the contribution of geometry and stylus-related properties to the classification and to obtain the best-performing model.


\subsubsection{Feature Significance Testing}
\label{sub_sub_sec:FEATURE_SIGNIFICANCE_TESTING}

To validate whether our justifications for features' definitions align with the data statistics and to ascertain their relevance in the intended classification task, we individually and independently evaluate the importance of each feature with respect to its significance for predicting the target variable. To this end, for each extracted feature in our collected dataset, its influence on the binary target variable (type of the stroke) is assessed using the univariate Mann-Whitney U test \cite{mcknight2010mann}, calculating its corresponding \(p\)-value. After obtaining the \(p\)-values for all features, they are subsequently assessed using the Benjamini Hochberg procedure \cite{benjamini2001control} to determine which features to retain and which to omit. The results are presented in Table \ref{tab:FEATURE_IMPORTANCES}, listing features in descending order based on their significance. As shown in the table, AvgPressure and AvgTilt are the only features considered to be irrelevant in our classification task. Excluding these features, we proceed to an ablation study as described in the following section.

\begin{table}[]
    \centering
    \begin{tabular}{c|c|c|c}
        Feature & $\mathcal{GEO}$ & $\mathcal{STY}$ & $p$-value\\
        \hline
        \hline
        \textit{PrimSegCount} & \ding{51} & \ding{51} & $1.83\mathrm{e}-161{<0.05}^{*}$\\
        \hline
        \textit{Density} & \ding{51} & \ding{51} & $7.05\mathrm{e}-129{<0.05}^{*}$\\
        \hline
        \textit{Straightness} & \ding{51} & \ding{55} & $7.99\mathrm{e}-83{<0.05}^{*}$\\
        \hline
        \textit{Length} & \ding{51} & \ding{55} & $3.25\mathrm{e}-48{<0.05}^{*}$\\
        \hline
        \textit{Order} & \ding{55} & \ding{51} & $6.92\mathrm{e}-41{<0.05}^{*}$\\
        \hline
        \textit{Duration} & \ding{55} & \ding{51} & $3.47\mathrm{e}-40{<0.05}^{*}$\\
        \hline
        \textit{Dist} & \ding{51} & \ding{55} & $7.69\mathrm{e}-23{<0.05}^{*}$\\
        \hline
        \textit{ColorShift} & \ding{55} & \ding{51} & $3.22\mathrm{e}-7{<0.05}^{*}$\\
        \hline
        \textit{AvgSpeed} & \ding{55} & \ding{51} & $2.26\mathrm{e}-5{<0.05}^{*}$\\
        \hline
        \textit{AvgPressure} & \ding{55} & \ding{51} & $3.44\mathrm{e}-1$\\
        \hline
        \textit{AvgTilt} & \ding{55} & \ding{51} & $7.04\mathrm{e}-1$\\
        \hline
    \end{tabular}
    \begin{tablenotes}
        \item [*] * $p$-values less than 0.05 considered to be significant.
    \end{tablenotes}
    \caption{Relevance table for extracted features with respect to the binary target variable (Shape versus Scribble). The $\mathcal{GEO}$ ($\mathcal{STY}$) column is checked if the feature's computation, according to its definition, involves geometry- (stylus-) related properties from the interface-recorded metadata.}
    \label{tab:FEATURE_IMPORTANCES}
\end{table}


\subsubsection{Ablation Studies}
\label{sub_sub_sec:ABLATION STUDIES}

There are several features associated with our classification task. To further understand how geometry-related and stylus-related properties convey useful information related to the design intent in our classification task, we run ablation tests with two meta-estimators, Random Forest (RF) \cite{breiman2001random} and XGBoost \cite{chen2016xgboost} (XGBRF) classifiers, on four different subset of retained features from Table \ref{tab:FEATURE_IMPORTANCES} as defined below: 
\begin{itemize}
    \item $\mathcal{GEO}$: Features only involving geometry-related properties in their computation and no stylus-related properties. 
    \item $\mathcal{STY}$: Features only involving stylus-related properties in their computation and no geometry-related properties. 
    \item $\mathcal{GEO}\vee\mathcal{STY}$: All retained features that incorporate either geometry or stylus-related properties in their computation. 
    \item $\mathcal{GEO}\wedge\mathcal{STY}$: Features that require at least a geometry and at least a stylus-related property in their computation. 
\end{itemize}

The classifiers are trained and tested using an $80\%$-$20\%$ data split from our collected dataset (Section \ref{sub_sec:DATASET}). We perform robust standardization on the features using the $5\textsuperscript{th}$ and $95\textsuperscript{th}$ percentiles of each feature. Also, for training, a grid search is performed over specified parameter values for the estimator to select optimal hyper-parameters using $5$-fold cross-validation. We then refit an estimator using the resultant optimal hyper-parameters on the training set and report its scores on the unseen fresh test set, as outlined in Table \ref{tab:ABLATION_RESULTS}. According to reported metrics, the best model is $\text{RF}_{\mathcal{GEO}\vee\mathcal{STY}}$. Compared with other models, the superiority of this model in terms of the reported metrics suggests that both geometry and stylus-related properties are vital to the performance of the model in distinguishing the type of drawn strokes.  

\begin{table}[]
    \centering
    \begin{tabular}{c|c|c|c|c}
        Model & Set & Accuracy & Precision & Recall\\
        \hline
        \hline
         \multirow{2}{7.125em}{\small $\text{RF}_{\mathcal{GEO}\vee\mathcal{STY}}$} & Train & 99.64 & 99.39 & 100.00\\
         & Test & \textbf{99.52} & \textbf{99.14} & \textbf{100.00}\\
         \hline
         \multirow{2}{7.125em}{\small $\text{RF}_{\mathcal{GEO}}$} & Train & 99.40 & 99.59 & 99.39\\
         & Test & 89.95 & 90.59 & 91.37\\
         \hline
         \multirow{2}{7.125em}{\small $\text{RF}_{\mathcal{STY}}$} & Train & 99.88 & 99.79 & 100.00\\
         & Test & 96.65 & 94.30 & \textbf{100.00}\\
         \hline
         \multirow{2}{7.125em}{\small $\text{RF}_{\mathcal{GEO}\wedge\mathcal{STY}}$} & Train & 98.92 & 98.39 & 99.79\\
         & Test & 99.04 & 98.30 & \textbf{100.00}\\
         \hline
         \multirow{2}{7.125em}{\small $\text{XGBRF}_{\mathcal{GEO}\vee\mathcal{STY}}$} & Train & 99.16 & 99.38 & 99.18\\
         & Test & 98.56 & 97.47 & \textbf{100.00}\\
         \hline
         \multirow{2}{7.125em}{\small $\text{XGBRF}_{\mathcal{GEO}}$} & Train & 91.86 & 92.74 & 93.49\\
         & Test & 89.95 & 89.25 & 93.10\\
         \hline
         \multirow{2}{7.125em}{\small $\text{XGBRF}_{\mathcal{STY}}$} & Train & 98.44 & 97.61 & 99.79\\
         & Test & 95.69 & 93.49 & 99.13\\
         \hline
         \multirow{2}{7.125em}{\small $\text{XGBRF}_{\mathcal{GEO}\wedge\mathcal{STY}}$} & Train & 98.80 & 98.58 & 99.39\\
         & Test & 99.04 & 98.30 & \textbf{100.0}\\
         \hline
    \end{tabular}
    \caption{Ablation results for two models, evaluated across four feature subsets as indicated by the model's subscripts.}
    \label{tab:ABLATION_RESULTS}
\end{table}


\subsection{Our Reconstructions}
\label{sub_sec:OUR_RESULTS}

\begin{figure*}[]
  \centering
  \includegraphics[width=1.0\linewidth]{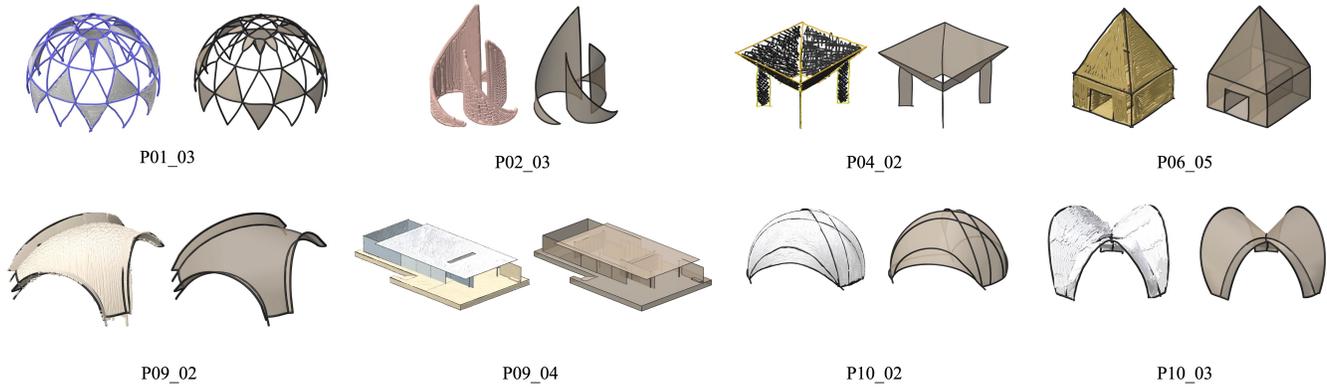}
  \caption{A number of input sketches and the reconstructions produced by Strokes2Surface. Each is captioned by its corresponding Participant ID who sketched the object.}
  \label{fig:RESULTS_AND_VALIDATION}
\end{figure*}

We tested our pipeline on a a large number of inputs. These include sketches created by the participants in our user study and a few others drawn by the two architectural design modeling experts from our research group. These input sketches depict architectural objects of varying scales and complexities, ranging from small-scale theater stage design walls to large-scale pavilions. In most of the sketches, our outputs accurately reflect the designer-intended geometry (Figure \ref{fig:RESULTS_AND_VALIDATION}), confirming its robustness to scale and complexity of the sketched objects. We detail in Table \ref{tab:RUNTIMES} the scale, the complexity of sketches in terms of number of strokes they constitute, and runtime of the four main steps of our pipeline (Sections \ref{sub_sub_sec:CLUSTERING_SHAPE_STROKES}, \ref{sub_sub_sec:TOPOLOGY_RECOVERY}, \ref{sub_sub_sec:CLUSTERING_SCRIBBLE_STROKES}, and \ref{sub_sub_sec:CYCLE DISCOVERY_AND_GEOMETRY_GENERATION}) on the eight sketches shown in Figure \ref{fig:RESULTS_AND_VALIDATION} (refer to Supplementary Material for additional results). Timings vary from a few seconds on small-scale sketches with a low amount of strokes up to a few minutes (max. $5$ minutes) on large-scale sketches with a very high number of strokes. The bottlenecks reside mainly in the feature extraction step in the very beginning for the classifier's inference and in the topology recovery step when the sketch is complex and possesses a large number of strokes. On the other hand, clustering Shape and Scribble strokes takes less than $5$ and $12$ seconds in most cases, respectively.  

\begin{table}[]
    \centering
    \begin{tabular}{c | c | c | c | c | c | c} 
        \small ID & \small $\text{Scale}^{*}$ & \small Strokes & (i) & (ii) & (iii) & (iv) \\
        \hline
        \hline
        \small P01\_03 & 7.2 & 198 & 25.1s & 4.7s & 108.8s & 5.6s\\ 
        \hline
        \small P02\_03 & 3.2 & 28 & 4.4s & 1.5s & 1.5s & 1.4s\\ 
        \hline
        \small P04\_02 & 12.1 & 54 & 9.6s & 2.6s & 11.7s & 1.4s\\ 
        \hline
        \small P06\_05 & 4.3 & 62 & 12.5s & 1.9s & 11.4s & 3.3s\\
        \hline
        \small P09\_02 & 15.1 & 306 & 107.6s & 5.7s & 5.1s & 185.8s\\
        \hline
        \small P09\_04 & 36.1 & 402 & 284.5s & 37.8s & 159.9s & 101.1s\\
        \hline
        \small P10\_02 & 15.1 & 50 & 7.3s & 6.2s & 11.0s & 1.9s\\
        \hline
        \small P10\_03 & 8.1 & 70 & 13.9s & 2.5s & 3.9s & 2.8s\\
        \hline
    \end{tabular}
    \begin{tablenotes}
        \item [*] * Scale is computed as the diameter of the sketch's axis-aligned bounding box.
    \end{tablenotes}
    \caption{For each sketch in Figure \ref{fig:RESULTS_AND_VALIDATION}, we provide the scale, complexity in terms of total number of strokes, and runtimes for the main four steps of the pipeline: (i) Stroke Type Prediction, (ii) Clustering Shape Strokes, (iii) Topology Recovery, and (iv) Clustering Scribble Strokes. All the computations were carried out on an Apple M1 chip featuring an 8-core CPU with 16GB memory.}
    \label{tab:RUNTIMES}
\end{table}


\subsubsection{Comparison to Prior Art}
\label{sub_sub_sec:COMPARISON_TO_PRIOR_ART}

Strokes2Surface recovers curve networks from a dense set of 3D strokes, coupled with timestamp (fourth dimension) and additional metadata recorded by the interface at the time of sketching. Therefore, its inputs do not exhibit exactly the same characteristics as the ones created by other 3D interfaces. However, aside from the fact that our input sketches can be easily converted into point clouds, the output produced by clustering Shape strokes, once consolidated, aligns with the input specifications of stroke cloud surfacing methods and can be compared with those. Furthermore, to highlight the effect of the clustering Scribble strokes in cycle discovery, we present comparisons of our recovered curve networks surfaced with and without taking this step into account. In the following sections, we provide targeted comparisons of our reconstructions against state-of-the-art methods from point cloud, stroke cloud, and curve network surfacing methods.

\begin{figure*}[]
    \centering
    \includegraphics[width=1.005\linewidth]{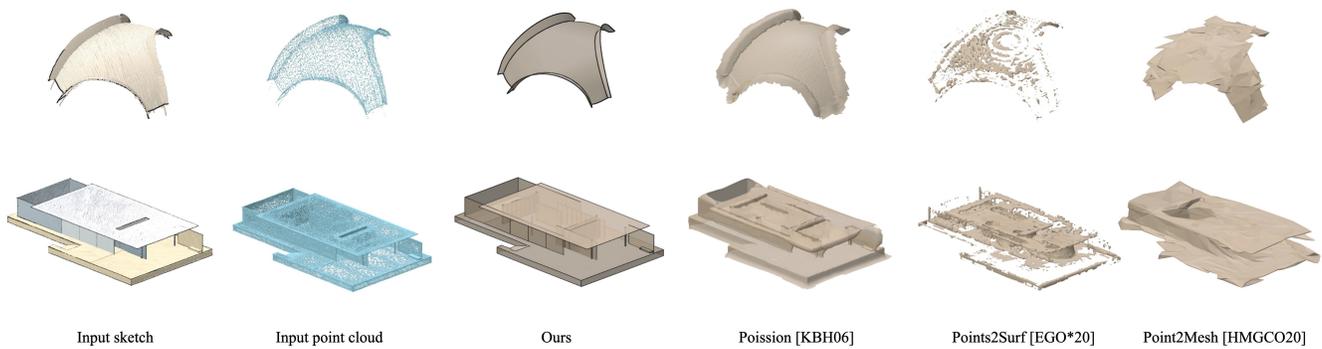}
    \caption{Comparison of our reconstructions against point cloud reconstruction techniques using input strokes polyline vertices. For the methods requiring per point normals, we provided the stroke's parent canvas normal for each vertex of the stroke polyline.}
    \label{fig:COMPARISON_SURFACING_POINT_CLOUDS}
\end{figure*}

\paragraph*{Comparison Against Point Cloud Surfacing.} Our input sketches can be easily converted into point clouds using stroke polyline vertices or by sampling points on the triangle strips or square sweeps forming the sketch's constituent strokes. Most of the existing point cloud reconstruction methods by design require dense point clouds and are doomed to fail on 3D sketches that only provide very sparse and non-uniform sampled input data. Figure \ref{fig:COMPARISON_SURFACING_POINT_CLOUDS} shows comparisons of our outputs to those produced from Poisson \cite{kazhdan2006poisson}, and two recent learning-based point cloud reconstruction methods, Points2Surf \cite{erler2020points2surf} and Point2Mesh \cite{hanocka2020point2mesh}. For the Poisson reconstruction, we set the $depth$ of the octree used for the surface reconstruction to $8$, and for Points2Surf, we used their provided pre-trained best model based on their ablation results. Point2Mesh learns from a single object by optimizing the weights of a CNN to deform an initial mesh to shrink-wrap the input point cloud. Following what they suggested in their implementation, we computed the convex hull of the input sketch point cloud, used it as the initial mesh, and ran the algorithm for $6000$ iterations. As evident in Figure \ref{fig:COMPARISON_SURFACING_POINT_CLOUDS}, these methods catastrophically fail on 3D sketches, producing meshes with multiple redundant connected components and mesh triangles connecting unrelated surface parts. Thanks to the underlying curve network structure, our method yields a collection of surface patches joined together on the curve segments rather than a single triangle mesh. The availability of individual surface patches could be beneficial for several architectural design applications within which each patch could represent an architectural element, e.g., a roof or a wall in Building Information Modeling (BIM) context. Also, the curve network segments themselves could be exploited for various architectural design purposes, e.g., paneling surfaces \cite{eigensatz2010paneling} or simplification for maintaining structural stability \cite{neveu2022stability}.

\begin{figure}[]
    \centering
    \includegraphics[width=1.\linewidth]{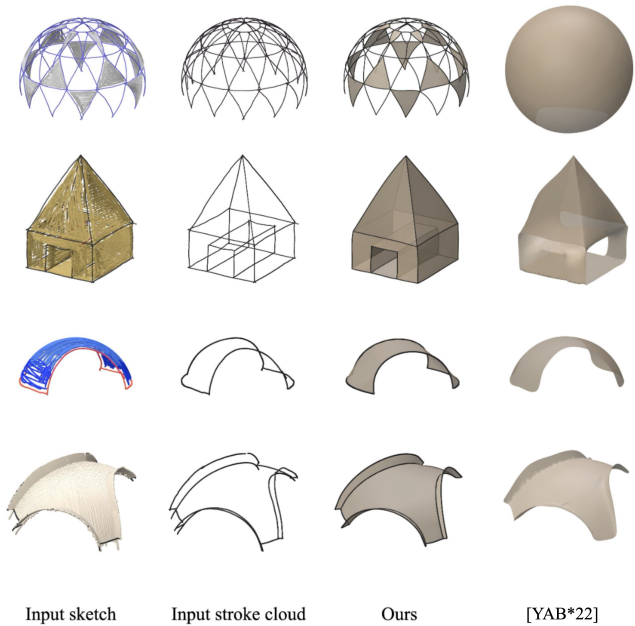}
    \caption{Comparison of our reconstructions against \cite{yu2022piecewise}. Consolidated clusters of Shape strokes are given as the input stroke cloud to their method.}
    \label{fig:COMPARISON_SURFACING_STROKE_CLOUDS}
\end{figure}

\paragraph*{Comparison Against Stroke Cloud Surfacing.} After the strokes are classified to determine their type, by omitting those identified as Scribble, what remains -- when consolidating those classified as Shape before topology recovery -- is a set of strokes that depict the boundaries of the intended object but without their connectivities fixed. This can be fed as input to stroke cloud surfacing methods. To this end, given this stroke cloud as input to the method described in \cite{yu2022piecewise}, we compare our reconstructions with theirs, as depicted in Figure \ref{fig:COMPARISON_SURFACING_STROKE_CLOUDS}. Their method requires an approximate \textit{proxy surface} to which they project the strokes and further segment the surface into smooth patches joined sharply along some strokes, and optimize these patches to fit surrounding strokes. Following what their implementation suggests, we used the VIPSS algorithm \cite{huang2019variational} to compute the initial proxy mesh from the consolidated Shape strokes and then manually annotated them as boundary curves before feeding them to their method. In cases where the design has a manifold shape, their method reconstructs very similar results to ours (Figure \ref{fig:COMPARISON_SURFACING_STROKE_CLOUDS}, third and fourth rows). However, it encounters errors in cases where the designed architectural object is non-manifold and contains openings or holes (Figure \ref{fig:COMPARISON_SURFACING_STROKE_CLOUDS}, first and second rows) and ends up connecting unintended parts of the geometry together or leaving some void areas. Compared to their results, our reconstructions are vulnerable in reconstructing surfaces with high curvature when there are only a few surrounding strokes present. This limitation can potentially mitigated by using the boundary normals when triangulating the cycle patches.

\begin{figure*}[]
    \centering
    \includegraphics[width=0.875\linewidth]{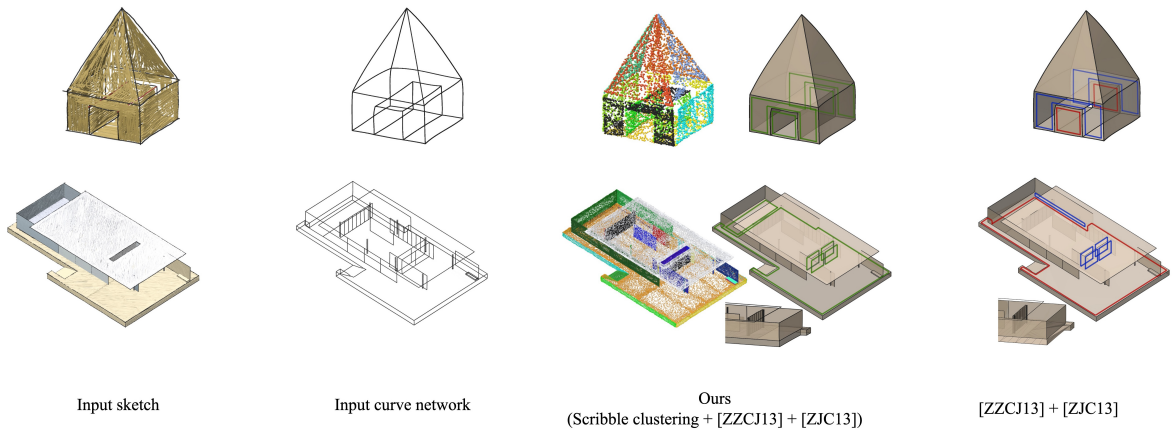}
    \caption{Comparison of our reconstructions \textit{with} and \textit{without} using Scribble clusters for guiding the cycle discovery using the \cite{zhuang2013general} and \cite{zou2013algorithm}. The wrong cycles detected are shown in red, and the missed ones are shown in blue, while the corresponding cycles detected by ours are shown in green.}
    \label{fig:COMPARISON_SURFACING_CURVE_NETWORKS}
\end{figure*}

\paragraph*{Comparison Against Curve Network Surfacing.} Presence of Scribble strokes is not required by the pipeline, rather an often observed practice designers employ. However, to evaluate their impact on cycle discovery in the final reconstructions, we applied the method from \cite{zhuang2013general} in two ways: first, to the subset of curves in our recovered curve networks determined by bounding box of the Scribble clusters; and second, to the complete curve network without considering the Scribble clusters. In both cases, we surface the identified cycles using the method described in \cite{zou2013algorithm} (Figure \ref{fig:COMPARISON_SURFACING_CURVE_NETWORKS}). For our tests, we set all curves' \textit{capacity} values to $2$, in line with the default assumptions in their paper. As depicted in Figure \ref{fig:COMPARISON_SURFACING_CURVE_NETWORKS}, when cycles are searched for globally within the whole curve network rather than locally leveraging Scribbles, and the geometry is non-manifold or has holes -- a scenario frequently observed in architectural design -- their algorithm struggles to correctly identify all cycles. At times, due to the non-manifoldness of the underlying object, it skips some cycles. This issue is evident on both the front and back sides of the geometry in the example shown in the first row of Figure \ref{fig:COMPARISON_SURFACING_CURVE_NETWORKS}. Furthermore, without verifying the existence of Scribble strokes against the surfaced bounding patches of cycles, some unintended areas are surfaced, such as the openings in the first row of Figure \ref{fig:COMPARISON_SURFACING_CURVE_NETWORKS}. As pointed out in their paper, some of these drawbacks could be addressed if the program could automatically identify possibly non-manifold curves and suggest their capacity.


\section{Conclusion}
\label{sec:CONCLUSION}

We presented Strokes2Surface, an offline geometry reconstruction pipeline for 4D architectural design sketches, aimed at bridging the gap between concept design and digital modeling stages. The pipeline is supported by three machine learning models: A binary classifier responsible for stroke type recognition and the two density-based clustering models responsible for parsing strokes of each type into groups representing boundaries and edges, and enclosing areas and faces, enabling recovery of a curve network from the design sketch. To the best of our knowledge, this is the first \textit{offline} geometry reconstruction of 3D/4D sketches in the architectural design domain. Furthermore, the pipeline introduces the following key standalone technical contributions:
\begin{itemize}
    \item The Shape/Scribble dichotomy, based on architectural sketching practices, and our hand-engineered features as input to the classifier are introduced for the first time in the 3D sketching domain. Our stroke classifier has the capability to be used in existing pen-on-tablet sketching interfaces that allow creation of 3D sketches, i.e., MentalCanvas \cite{dorsey2007mental} and NapkinSketch \cite{xin2008napkin}.
    \item Our pipeline introduces a 3D sketch consolidation method that handles imprecise sketches from various 3D sketching interfaces and is not limited to pen-on-tablet interfaces, as it only relies on 3D stroke vertices and brush thickness values.
    \item Our Stroke Classifier and Shape Clustering models' outputs can be combined to provide suitable input for stroke cloud surfacing methods, thereby facilitating 3D reconstruction in pen-on-tablet interfaces not inherently coupled with a modeling algorithm.
    \item We presented the first dataset of 3D/4D architectural design sketches in the literature for further use in the community. 
\end{itemize}

\paragraph*{Limitations and Future Work.} Despite our method's contributions, there is still potential for further improvements in various directions. As mentioned in Section \ref{sec:INPUT_DRAWING_CHARACTERISTICS}, scribbling is one of the few techniques for drawing the surfaces of forms and shapes. Extending our classifier to recognize strokes drawn using other drawing techniques would diversify the applicability of the pipeline to other sketching styles. In its current state, Strokes2Surface does not take into account the geometry of underlying strokes' parent canvases when surfacing the recovered curve networks, however, they could provide a powerful prior for more accurate reconstruction, i.e., to obtain more realistic shape curvature. What we further plan to address in our future work is taking into account the spatial relationship among different networks recovered from an input design sketch to avoid scenarios where a surface patch ends up covering another area which is supposed be an opening and void within another patch (see Figure \ref{fig:RESULTS_AND_VALIDATION} (P09\_04)). Furthermore, obtaining parametric geometries is crucial for many downstream design pipelines, for which we find extending \cite{li2022free2cad} to 3D sketches an intriguing direction for future research.


\section*{Acknowledgments}
\label{sec:ACKNOWLEDGMENTS}

This research was funded by Austrian Science Fund (FWF) project F 77 (SFB “Advanced Computational Design”). We would like to thank our partners in ACD Sub-Project 2 who provided valuable input for the topics of this paper: Michael Hensel, Peter Ferschin, Julia Resiginer, Ingrid Erb, Balint Istvan Kovacs, and Dalel Daleyev.

\bibliographystyle{eg-alpha-doi}  
\bibliography{main}        

\newcommand{\etalchar}[1]{$^{#1}$}
\begin{thebibliography}{\uppercase{HMGCO20}}

\bibitem[AJA12]{abbasinejad2012surface}
\textsc{Abbasinejad F., Joshi P., Amenta N.}:
\newblock Surface patches from unorganized space curves.
\newblock In \emph{Proceedings of the twenty-eighth annual symposium on Computational geometry} (2012), pp.~417--418.

\bibitem[BAOBK12]{batuhan2012free}
\textsc{Batuhan~Arisoy E., Orbay G., Burak~Kara L.}:
\newblock Free form surface skinning of 3d curve clouds for conceptual shape design.
\newblock \emph{Journal of computing and information science in engineering 12}, 3 (2012), 031005.

\bibitem[Bre01]{breiman2001random}
\textsc{Breiman L.}:
\newblock Random forests.
\newblock \emph{Machine learning 45} (2001), 5--32.

\bibitem[BTS05]{barla2005geometric}
\textsc{Barla P., Thollot J., Sillion F.~X.}:
\newblock Geometric clustering for line drawing simplification.
\newblock In \emph{ACM SIGGRAPH 2005 Sketches}. 2005, pp.~96--es.

\bibitem[BY01]{benjamini2001control}
\textsc{Benjamini Y., Yekutieli D.}:
\newblock The control of the false discovery rate in multiple testing under dependency.
\newblock \emph{Annals of statistics} (2001), 1165--1188.

\bibitem[CG16]{chen2016xgboost}
\textsc{Chen T., Guestrin C.}:
\newblock Xgboost: A scalable tree boosting system.
\newblock In \emph{Proceedings of the 22nd acm sigkdd international conference on knowledge discovery and data mining} (2016), pp.~785--794.

\bibitem[Chi19]{ching2019design}
\textsc{Ching F.~D.}:
\newblock \emph{Design drawing}.
\newblock John Wiley \& Sons, 2019.

\bibitem[CKX{\etalchar{*}}08]{chen2008sketching}
\textsc{Chen X., Kang S.~B., Xu Y.-Q., Dorsey J., Shum H.-Y.}:
\newblock Sketching reality: Realistic interpretation of architectural designs.
\newblock \emph{ACM Transactions on Graphics (TOG) 27}, 2 (2008), 1--15.

\bibitem[CLHC14]{chien2014line}
\textsc{Chien Y., Lin W.-C., Huang T.-S., Chuang J.-H.}:
\newblock Line drawing simplification by stroke translation and combination.
\newblock In \emph{Fifth International Conference on Graphic and Image Processing (ICGIP 2013)} (2014), vol.~9069, SPIE, pp.~180--185.

\bibitem[DA22]{dzurillas}
\textsc{Dzurilla D., Achten H.}:
\newblock What’s happening to architectural sketching?

\bibitem[DLP{\etalchar{*}}22]{deng2022sketch2pq}
\textsc{Deng Z., Liu Y., Pan H., Jabi W., Zhang J., Deng B.}:
\newblock Sketch2pq: freeform planar quadrilateral mesh design via a single sketch.
\newblock \emph{IEEE Transactions on Visualization and Computer Graphics} (2022).

\bibitem[Do02]{do2002drawing}
\textsc{Do E. Y.-L.}:
\newblock Drawing marks, acts, and reacts: Toward a computational sketching interface for architectural design.
\newblock \emph{AI EDAM 16}, 3 (2002), 149--171.

\bibitem[DP73]{douglas1973algorithms}
\textsc{Douglas D.~H., Peucker T.~K.}:
\newblock Algorithms for the reduction of the number of points required to represent a digitized line or its caricature.
\newblock \emph{Cartographica: the international journal for geographic information and geovisualization 10}, 2 (1973), 112--122.

\bibitem[DXS{\etalchar{*}}07]{dorsey2007mental}
\textsc{Dorsey J., Xu S., Smedresman G., Rushmeier H., McMillan L.}:
\newblock The mental canvas: A tool for conceptual architectural design and analysis.
\newblock In \emph{15th Pacific Conference on Computer Graphics and Applications (PG'07)} (2007), IEEE, pp.~201--210.

\bibitem[EGO{\etalchar{*}}20]{erler2020points2surf}
\textsc{Erler P., Guerrero P., Ohrhallinger S., Mitra N.~J., Wimmer M.}:
\newblock Points2surf learning implicit surfaces from point clouds.
\newblock In \emph{European Conference on Computer Vision} (2020), Springer, pp.~108--124.

\bibitem[EKS{\etalchar{*}}96]{ester1996density}
\textsc{Ester M., Kriegel H.-P., Sander J., Xu X., et~al.}:
\newblock A density-based algorithm for discovering clusters in large spatial databases with noise.
\newblock In \emph{kdd} (1996), vol.~96, pp.~226--231.

\bibitem[EKS{\etalchar{*}}10]{eigensatz2010paneling}
\textsc{Eigensatz M., Kilian M., Schiftner A., Mitra N.~J., Pottmann H., Pauly M.}:
\newblock Paneling architectural freeform surfaces.
\newblock In \emph{ACM SIGGRAPH 2010 papers}. 2010, pp.~1--10.

\bibitem[FK10]{fabbri20103d}
\textsc{Fabbri R., Kimia B.}:
\newblock 3d curve sketch: Flexible curve-based stereo reconstruction and calibration.
\newblock In \emph{2010 IEEE Computer Society Conference on Computer Vision and Pattern Recognition} (2010), IEEE, pp.~1538--1545.

\bibitem[FMRU03]{fiorentino20033d}
\textsc{Fiorentino M., Monno G., Renzulli P.~A., Uva A.~E.}:
\newblock 3d sketch stroke segmentation and fitting in virtual reality.
\newblock In \emph{International conference on the Computer Graphics and Vision} (2003), vol.~5, Citeseer.

\bibitem[GHL{\etalchar{*}}20]{gryaditskaya2020lifting}
\textsc{Gryaditskaya Y., H{\"a}hnlein F., Liu C., Sheffer A., Bousseau A.}:
\newblock Lifting freehand concept sketches into 3d.
\newblock \emph{ACM Transactions on Graphics (TOG) 39}, 6 (2020), 1--16.

\bibitem[HCJ19]{huang2019variational}
\textsc{Huang Z., Carr N., Ju T.}:
\newblock Variational implicit point set surfaces.
\newblock \emph{ACM Transactions on Graphics (TOG) 38}, 4 (2019), 1--13.

\bibitem[HGSB22]{hahnlein2022symmetry}
\textsc{H{\"a}hnlein F., Gryaditskaya Y., Sheffer A., Bousseau A.}:
\newblock Symmetry-driven 3d reconstruction from concept sketches.
\newblock In \emph{ACM SIGGRAPH 2022 Conference Proceedings} (2022), pp.~1--8.

\bibitem[HMGCO20]{hanocka2020point2mesh}
\textsc{Hanocka R., Metzer G., Giryes R., Cohen-Or D.}:
\newblock Point2mesh: A self-prior for deformable meshes.
\newblock \emph{arXiv preprint arXiv:2005.11084} (2020).

\bibitem[KBH06]{kazhdan2006poisson}
\textsc{Kazhdan M., Bolitho M., Hoppe H.}:
\newblock Poisson surface reconstruction.
\newblock In \emph{Proceedings of the fourth Eurographics symposium on Geometry processing} (2006), vol.~7, p.~0.

\bibitem[KEKF23]{kovacs2023mr}
\textsc{Kovacs B.~I., Erb I., Kaufmann H., Ferschin P.}:
\newblock Mr. sketch. immediate 3d sketching via mixed reality drawing canvases.
\newblock In \emph{2023 IEEE International Symposium on Mixed and Augmented Reality (ISMAR)} (2023), IEEE, pp.~10--19.

\bibitem[KH13]{kazhdan2013screened}
\textsc{Kazhdan M., Hoppe H.}:
\newblock Screened poisson surface reconstruction.
\newblock \emph{ACM Transactions on Graphics (ToG) 32}, 3 (2013), 1--13.

\bibitem[LABS23]{liu2023stripmaker}
\textsc{Liu C., Aoki T., Bessmeltsev M., Sheffer A.}:
\newblock Stripmaker: Perception-driven learned vector sketch consolidation.
\newblock \emph{ACM Transactions on Graphics (TOG) 42}, 4 (2023), 1--15.

\bibitem[LCX{\etalchar{*}}23]{luo20233d}
\textsc{Luo L., Chowdhury P.~N., Xiang T., Song Y.-Z., Gryaditskaya Y.}:
\newblock 3d vr sketch guided 3d shape prototyping and exploration.
\newblock In \emph{Proceedings of the IEEE/CVF International Conference on Computer Vision} (2023), pp.~9267--9276.

\bibitem[Lee00]{lee2000curve}
\textsc{Lee I.-K.}:
\newblock Curve reconstruction from unorganized points.
\newblock \emph{Computer aided geometric design 17}, 2 (2000), 161--177.

\bibitem[LKB22]{lee2022rapid}
\textsc{Lee J.~H., Kim H., Bae S.-H.}:
\newblock Rapid design of articulated objects.
\newblock \emph{ACM Transactions on Graphics (TOG) 41}, 4 (2022), 1--8.

\bibitem[LPBM22]{li2022free2cad}
\textsc{Li C., Pan H., Bousseau A., Mitra N.~J.}:
\newblock Free2cad: Parsing freehand drawings into cad commands.
\newblock \emph{ACM Transactions on Graphics (TOG) 41}, 4 (2022), 1--16.

\bibitem[LRS18]{liu2018strokeaggregator}
\textsc{Liu C., Rosales E., Sheffer A.}:
\newblock Strokeaggregator: Consolidating raw sketches into artist-intended curve drawings.
\newblock \emph{ACM Transactions on Graphics (TOG) 37}, 4 (2018), 1--15.

\bibitem[Mah18]{mahoney2018v}
\textsc{Mahoney J.~M.}:
\newblock The v-sketch system, machine assisted design exploration in virtual reality.

\bibitem[MN10]{mcknight2010mann}
\textsc{McKnight P.~E., Najab J.}:
\newblock Mann-whitney u test.
\newblock \emph{The Corsini encyclopedia of psychology} (2010), 1--1.

\bibitem[NPTB22]{neveu2022stability}
\textsc{Neveu W., Puhachov I., Thomaszewski B., Bessmeltsev M.}:
\newblock Stability-aware simplification of curve networks.
\newblock In \emph{ACM SIGGRAPH 2022 Conference Proceedings} (2022), pp.~1--9.

\bibitem[OK11]{orbay2011beautification}
\textsc{Orbay G., Kara L.~B.}:
\newblock Beautification of design sketches using trainable stroke clustering and curve fitting.
\newblock \emph{IEEE Transactions on Visualization and Computer Graphics 17}, 5 (2011), 694--708.

\bibitem[OMYA16]{ogawa2016sketch}
\textsc{Ogawa T., Matsui Y., Yamasaki T., Aizawa K.}:
\newblock Sketch simplification by classifying strokes.
\newblock In \emph{2016 23rd International Conference on Pattern Recognition (ICPR)} (2016), IEEE, pp.~1065--1070.

\bibitem[PLS{\etalchar{*}}15]{pan2015flow}
\textsc{Pan H., Liu Y., Sheffer A., Vining N., Li C.-J., Wang W.}:
\newblock Flow aligned surfacing of curve networks.
\newblock \emph{ACM Transactions on Graphics (TOG) 34}, 4 (2015), 1--10.

\bibitem[RRS19]{rosales2019surfacebrush}
\textsc{Rosales E., Rodriguez J., Sheffer A.}:
\newblock Surfacebrush: from virtual reality drawings to manifold surfaces.
\newblock \emph{arXiv preprint arXiv:1904.12297} (2019).

\bibitem[S{\etalchar{*}}19]{polyscope}
\textsc{Sharp N., et~al.}:
\newblock Polyscope, 2019.
\newblock www.polyscope.run.

\bibitem[SHBSS16]{stanko2016smooth}
\textsc{Stanko T., Hahmann S., Bonneau G.-P., Saguin-Sprynski N.}:
\newblock Smooth interpolation of curve networks with surface normals.
\newblock In \emph{Eurographics 2016 Short Papers} (2016), Eurographics Association, pp.~21--24.

\bibitem[TF22]{tono2022vitruvio}
\textsc{Tono A., Fischer M.}:
\newblock Vitruvio: 3d building meshes via single perspective sketches.
\newblock \emph{arXiv preprint arXiv:2210.13634} (2022).

\bibitem[VMLV{\etalchar{*}}21]{van2021strokestrip}
\textsc{Van~Mossel D.~P., Liu C., Vining N., Bessmeltsev M., Sheffer A.}:
\newblock Strokestrip: Joint parameterization and fitting of stroke clusters.
\newblock \emph{ACM Transactions on Graphics (TOG) 40}, 4 (2021), 1--18.

\bibitem[XCS{\etalchar{*}}14]{xu2014true2form}
\textsc{Xu B., Chang W., Sheffer A., Bousseau A., McCrae J., Singh K.}:
\newblock True2form: 3d curve networks from 2d sketches via selective regularization.
\newblock \emph{ACM Transactions on Graphics 33}, 4 (2014).

\bibitem[XSS08]{xin2008napkin}
\textsc{Xin M., Sharlin E., Sousa M.~C.}:
\newblock Napkin sketch: handheld mixed reality 3d sketching.
\newblock In \emph{Proceedings of the 2008 ACM symposium on Virtual reality software and technology} (2008), pp.~223--226.

\bibitem[YAB{\etalchar{*}}22]{yu2022piecewise}
\textsc{Yu E., Arora R., Baerentzen J.~A., Singh K., Bousseau A.}:
\newblock Piecewise-smooth surface fitting onto unstructured 3d sketches.
\newblock \emph{ACM Transactions on Graphics (TOG) 41}, 4 (2022), 1--16.

\bibitem[YAS{\etalchar{*}}21]{yu2021cassie}
\textsc{Yu E., Arora R., Stanko T., B{\ae}rentzen J.~A., Singh K., Bousseau A.}:
\newblock Cassie: Curve and surface sketching in immersive environments.
\newblock In \emph{Proceedings of the 2021 CHI Conference on Human Factors in Computing Systems} (2021), pp.~1--14.

\bibitem[YDSG21]{yu2021scaffoldsketch}
\textsc{Yu X., DiVerdi S., Sharma A., Gingold Y.}:
\newblock Scaffoldsketch: Accurate industrial design drawing in vr.
\newblock In \emph{The 34th Annual ACM Symposium on User Interface Software and Technology} (2021), pp.~372--384.

\bibitem[ZJC13]{zou2013algorithm}
\textsc{Zou M., Ju T., Carr N.}:
\newblock An algorithm for triangulating multiple 3d polygons.
\newblock In \emph{Computer graphics forum} (2013), vol.~32, Wiley Online Library, pp.~157--166.

\bibitem[ZLDM16]{zheng2016smartcanvas}
\textsc{Zheng Y., Liu H., Dorsey J., Mitra N.~J.}:
\newblock Smartcanvas: Context-inferred interpretation of sketches for preparatory design studies.
\newblock In \emph{Computer Graphics Forum} (2016), vol.~35, Wiley Online Library, pp.~37--48.

\bibitem[ZZCJ13]{zhuang2013general}
\textsc{Zhuang Y., Zou M., Carr N., Ju T.}:
\newblock A general and efficient method for finding cycles in 3d curve networks.
\newblock \emph{ACM Transactions on Graphics (TOG) 32}, 6 (2013), 1--10.

\end{thebibliography}


\end{document}